\documentclass{article}
\usepackage[utf8]{inputenc}

\RequirePackage{fix-cm}
\usepackage{graphicx}
\usepackage{multirow,array}
\usepackage[labelfont=bf]{caption}
\usepackage{subcaption}
\usepackage{graphicx}
\usepackage{amssymb}
\usepackage{multicol}
\usepackage{booktabs}
\usepackage[usenames,dvipsnames]{xcolor}
\usepackage{framed} \definecolor{shadecolor}{cmyk}{0 0 0.2 0}
\usepackage[colorlinks=true,urlcolor = blue,citecolor=blue,linkcolor=blue]{hyperref}

\usepackage{diagbox}
\usepackage{natbib}

\usepackage{scalerel}
\usepackage{tikz}
\usepackage{amsmath}
\usetikzlibrary{svg.path}

\begin{document}

%%%%%%%%%%%%%%%%%%%%%%% file template.tex %%%%%%%%%%%%%%%%%%%%%%%%%
%
% This is a general template file for the LaTeX package SVJour3
% for Springer journals.          Springer Heidelberg 2010/09/16
%
% Copy it to a new file with a new name and use it as the basis
% for your article. Delete % signs as needed.
%
% This template includes a few options for different layouts and
% content for various journals. Please consult a previous issue of
% your journal as needed.
%
%%%%%%%%%%%%%%%%%%%%%%%%%%%%%%%%%%%%%%%%%%%%%%%%%%%%%%%%%%%%%%%%%%%
%
% First comes an example EPS file -- just ignore it and
% proceed on the \documentclass line
% your LaTeX will extract the file if required
%\begin{filecontents*}{example.eps}
%!PS-Adobe-3.0 EPSF-3.0
%%BoundingBox: 19 19 221 221
%%CreationDate: Mon Sep 29 1997
%%Creator: programmed by hand (JK)
%%EndComments
%gsave
%newpath
%  20 20 moveto
%  20 220 lineto
%  220 220 lineto
%  220 20 lineto
%closepath
%2 setlinewidth
%gsave
%  .4 setgray fill
%grestore
%stroke
%grestore
%\end{filecontents*}
%

% please place your own definitions here and don't use \def but
% \newcommand{}{}
%
% Insert the name of "your journal" with

\title{DB vs. OR}
\author{Mantas Radzvilas, Francesco De Pretis\footnote{Corresponding author: \href{mailto:francesco.depretis@unimore.it}{francesco.depretis@unimore.it}. Please, do not cite without permission.}, William Peden,\\ Daniele Tortoli and Barbara Osimani}
\date{November 2020}

\title{%PROVISIONAL TITLE:
Double blind vs. open review: an evolutionary game logit-simulating the behavior of authors and reviewers%\thanks{Grants or other notes
}


\maketitle

\begin{abstract}
Despite the tremendous successes of science in providing knowledge and technologies, the Replication Crisis has highlighted that scientific institutions have much room for improvement. Peer-review is one target of criticism and suggested reforms. However, despite numerous controversies peer review systems, plus the obvious complexity of the incentives affecting the decisions of authors and reviewers, there is very little systematic and strategic analysis of peer-review systems.

In this paper, we begin to address this feature of the peer-review literature by applying the tools of game theory. We use simulations to develop an evolutionary model based around a game played by authors and reviewers, before exploring some of its tendencies. In particular, we examine the relative impact of double-blind peer-review and open review on incentivising reviewer effort under a variety of parameters. We also compare (a) the impact of one review system versus another with (b) other alterations, such as higher costs of reviewing.

We find that is no reliable difference between peer-review systems in our model. Furthermore, under some conditions, higher payoffs for good reviewing can lead to less (rather than more) author effort under open review. Finally, compared to the other parameters that we vary, it is the exogenous utility of author effort that makes an important and reliable difference in our model, which raises the possibility that peer-review might not be an important target for institutional reforms.

\paragraph{Keywords} Double-Blind Peer Review; Open Review; Evolutionary Game Theory; Agent Based Modelling; Simulation; Replication Crisis.
\paragraph{Classifications} MSC 91A22; JEL C73.
\end{abstract}

\section{Introduction}
\label{intro}
The replication crisis has stimulated a colossal amount of methodological and institutional reflection in science. The core of the crisis is simple: research in a number of fields, like social psychology and medicine, replicates at a rate far lower than expected \citep{osc2015}. We all expect that some studies will not replicate, because inaccurate data is an unavoidable problem and unrepresentative samples will occur even with random sampling. Instead, the worry is that the rates of unreproducible research in some fields can only be explained by malpractice on the part of researchers. There is a large literature attempting to diagnose what has gone wrong, e.g. \citep{Ioannidis2005,Ioannidis2012,Simons2014,Fanelli2018,SEP2018}.

There are also many responses to the replication crisis. Some advocate reforms of statistical methodologies and concepts \citep{gelman2015,wasserstein2016,Mayo2018,trafimow2018}. Others, like the AllTrials campaign\footnote{\href{https://www.alltrials.net/}{https://www.alltrials.net/}} and the Center for Open Science\footnote{\href{https://cos.io/}{https://cos.io/}} propose greater transparency. Another approach is to change the broader institutional environment in which scientists operate\footnote{\href{https://blogs.canterbury.ac.nz/surejournal/}{https://blogs.canterbury.ac.nz/surejournal/}}. One general issue for reform proposals is that where there are incentives for bad scientific research, there might be an underlying tendency towards methodological degeneration that circumnavigates reforms that are targeted at particular malpractices \citep{Smaldino2016}. Changing the institutional environment could increase the incentives for good scientific practice and thereby provide a more stable and robust amelioration of the replication crisis.

As every scientist knows, a large part of their incentives can be summarised as ``publish or perish". Publication records are crucial factors in hiring decisions, grant approvals, conference invitations, and a scientist's general reputation within their discipline. However, it is normally impossible to read and systematically evaluate the entirety of a scientist's research when making these decisions. Scientists hence have an incentive to publish substandard work if they can, though this incentive can be counterbalanced by other considerations.

One counterbalance is peer review. Consequently, peer review is a natural institutional structure to contemplate reforming in order to reduce the bad practices behind the replication crisis. Some have argued that the institutions of peer review are net negatives and should be abolished \citep{heesen2019}. More typically, advocates of particular systems of peer review argue that their approach can improve the incentives of authors and reviewers, resulting in better research practices. There are many different types of peer review systems, which vary in multiple dimensions. A major dimension of difference is the blinding procedure. For example:

(1) Triple-blind: authors' identities are hidden to both reviewers and editors; reviewers' identities are hidden to authors. 

(2) Double-blind: the identities of authors and reviewers are hidden to each other.

(3) Single-blind: the identities of reviewers are hidden to authors, but not vice versa.

(4) Open review: no identities are hidden.\footnote{A journal might also blind with respect to results, in order to reduce the bias in favour of statistically significant results \citep{newcombe1987,Glymour2005}.}

In this article, we focus on double-blind peer review and open review. Advocates of both systems have presented a variety of arguments in their favour. Some argue that double-blinding leads to (a) less bias \citep{ferber1980,budden2008,lotfi2014,garvalov2015,darling2014,Bernard2018} and (b) better reviewer behaviour \citep{wendler2013}. Others contest the arguments for (a) \citep{fish1989,cox1993,armstrong1997,engqvist2008,webb2008,largent2016} and others argue that double-blinding impedes the vital role of reviewers in identifying conflicts of interest and redundant research \citep{pitkin1995}\footnote{However, see \citep{rogers2002}. One issue is that there are other means by which editors can accomplish these objectives \citep{peters1982,cox1993}. See also \href{https://retractionwatch.com/}{https://retractionwatch.com/}.}. Meanwhile, with respect to (b), advocates of open review also claim that this system leads to better reviewer behaviour \citep{turner2003,comer2014} and enables reviewers to gain credit for good reviewing practices \citep{chubin1990,godlee2002,eisenhart2002,raelin2008,wendler2013}. There is also a budding, but underdeveloped, empirical literature on the pros and cons of both systems, with no clear balance of evidence in favour of either approach\footnote{For an extensive survey of double-blind versus single-blind review, see \citep{snodgrass2006}.}. 

One challenge for empirical studies in this debate is that the journals and disciplines might have relevant differences. For instance, in a small subfield where the research agendas of other scientists are generally known, anonymity is unlikely to be effective. An additional challenge is that empirical regularities without a well-grounded analysis of the underlying incentives offer an unreliable foundation for policy changes, because the regularities might disappear if the incentive systems generating them are changed \citep{Lucas1976}. Yet the principal motivation for studying different peer review systems is exactly to decide whether to change peer review policies of journals, and thereby alter the incentives of scientists.

Our study helps address the issue of modelling the underlying incentives in peer-reviewed publishing. The informal debates regarding peer review systems and publication markets in general reveal that there are many incentives pushing authors and reviewers in different directions. We need systematic modelling in order to formalise the structure of these incentives, assess their dynamics, and think systematically about them. There is little research in this direction; for an exception, see \citep{Taylor2011}. Our aim is not to provide a definitive argument in favour of double-blinding or open review, but rather to develop and explore a game-theoretic model that will aid subsequent debates.

\section{Methods}

In this section, we describe our model and its underlying game. Our overall aim is a simple exploratory model of the incentives for authors and reviewers that allows comparisons of the effects of double-blind reviewing versus open reviewing on encouraging author effort. We focus on author effort rather than the quality of research because there is no consensus on how to quantify the latter or relate it to authors' decisions, whereas it is reasonable to model the former as a single parameter that is directly under authors' control.

We model choices made by authors and reviewers. To simplify, we assume that editors always follow reviewers' suggestions, as they do in the vast majority of decisions in the real world. We also assume that journal submissions are reviewed by a single reviewer. This assumption is not realistic, but it does correspond to what authors \emph{principally} care about, which is the aggregation of the reviewers' recommendations to accept or reject; since our ultimate interest is author effort, this assumption does not seem to hinder our main aims.

Our model generates evolutionary environments for author publication strategies. We explain it as follows: (1) the base game that schematises a particular interaction of an author and a reviewer and (2) our model that studies how author and reviewer strategies evolve over time under different initial conditions, i.e. parameters and institutional settings.

\subsection{The author-reviewer game}
The base game is a two player $6\times6$ normal form game\footnote{A normal form game is one where we completely specify the possible strategies and payoffs for a finite number of players.} which represents the one-shot interaction between player $a$ representing an author and player $r$ representing a reviewer. Player $a$'s set of strategies is a finite set of effort levels $E=\left(e_{1},...,e_{6}\right)$, where $e_{i}\in\left(0,1\right)$ for each $i\in\left(1,...,6\right)$.

Player $r$'s strategies is a set of acceptance thresholds $T=\left(t_{1},...,t_{6}\right)$, where $t_{i}\in\left(0,1\right)$ for each $i=\left(1,...,6\right)$. These correspond to a reviewer's decision upon considering whether a submitted article should be published. Player $r$ can either accept the paper (choose action $g$) or reject it (choose action $b$). Player $r$'s decision is determined by player $a$'s effort level and player $r$'s acceptance threshold. We define the acceptance function via a response function $d$ for a reviewer. Finally, let $g$ be the decision to accept and $b$ be the decision to reject. We can now define the reviewer's acceptance function as $d: E \times T\rightarrow\left\{g,b\right\}$, where $d\left(e_{i},t_{i}\right)=g$ for each $\left(e_{i},t_{i}\right)\in E\times T$, such that $e_{i}\geq t_{i}$; otherwise $d\left(e_{i},t_{i}\right)=b$. In words, the reviewer accepts the author's paper if it meets or exceeds the reviewer's requirements for effort, and rejects the paper if it does not.

We model Player $a$'s incentives in such a way that there are both costs and benefits to expending effort on their research. Player $a$'s base payoff from getting the paper accepted is 1, while rejection yields a base payoff 0. However, $a$ also receives additional payoff bonuses which depend on $a$'s effort level and $r$'s acceptance threshold. In case of acceptance, $a$ gets a payoff bonus $\epsilon e_{i}$, where $e_{i}>0$, which represents $a$'s reputation bonus. We assume this bonus to be proportional to $a$'s effort level. In addition, $a$ seeks to optimize the investment of effort and thus avoids investing more effort than is necessary to get the paper accepted. Thus, $a$ gets a payoff bonus $\alpha\left(1-e_{i}\right)$, where $\alpha>0$, from publishing a paper with less than maximum possible effort in every situation where the maximum effort is not required for acceptance. Player $a$'s effort is costly and the cost is proportional to the invested effort: higher effort is associated with a higher effort cost than lower effort. Player $a$'s effort cost will be defined as $\beta e$, where $\beta>0$.

By combining all the aforementioned incentives, we can define $a$'s final payoffs with a payoff function $\pi_{a}:E\times T\rightarrow\mathbb{R}$, which is such that 
\begin{equation}
\forall \left(e_{i},t_{i}\right)\in E\times T, \pi_{a}\left(e_{i},t_{i}\right)=\begin{cases}1+\alpha\left(1-e_{i}\right)+\epsilon e_{i}-\beta e_{i}\mbox{ when }e_{i}\geq t_{i}; \\
0-\beta e_{i} \mbox{ otherwise.} \end{cases}
\end{equation}

Thus, $a$ receives positive utility from acceptance, quantified by the positive terms at the top-right of the payoff function, from which their effort level is subtracted to give their overall utility outcome. In the case of rejection, they receive no boost, but suffer the same negative utility from their expenditure of effort.

Player $r$'s base payoff from accepting or rejecting $a$'s paper is 1. However, we also take $r$'s effort costs into account. To incorporate these costs into the model, $r$'s final payoff is affected by $a$'s effort and $r$'s own acceptance threshold. Just as it takes less effort to tell if a tall man is over 150 cm than if that man were approximately 150 cm, Player $r$ requires less effort to produce reports for papers which clearly are above $r$'s acceptance threshold than papers that are closer to it. Thus, $r$ gets a payoff bonus from reviewing easy-to-review papers. This bonus is defined as $\delta\left(e_{i}-t_{i}\right)$, where $\delta>0$.

By combining the aforementioned incentives, we can define $r$'s final payoffs with payoff function $\pi_{r}: E\times T\rightarrow\mathbb{R}$, which is such that 
\begin{equation}
\forall \left(e_{i},t_{i}\right)\in E\times T, \pi_{r}\left(e_{i},t_{i}\right)=\begin{cases}1+\delta\left(e_{i}-t_{i}\right)\mbox{ when }e_{i}\geq t_{i};\\
1\mbox{ otherwise}.
\end{cases}
\end{equation}

Thus, the parameter $\delta$ is a variable that determines the degree to which $r$'s utility is affected by the effort they save due to reviewing a paper that clearly exceeds their acceptance threshold.

We now consider the consequences of institutional environment on the structure of the game. Under open review, the community has access not only to $a$'s accepted work, but also $r$'s review of $a$'s accepted paper. In this case, the community can take a more active role in evaluating the quality of peer-review process and awarding a reputation bonus to authors and reviewers who promote higher than average effort standards. Under open review, $a$ receives a reputation bonus if it is accepted by $r$ who adopts a higher acceptance threshold than the population average effort level, while $r$ receives a reputation bonus if s/he accepts $a$'s paper which manifests a higher than average effort level. For $a$, this reputation bonus is defined as $\mu\mathrm{max}\left(t_{i}-\overline{e},0\right)$, where $\mu>0$ and $\overline{e}=\frac{1}{6}\sum_{i\in\left(1,...,6\right)}e_{i}$. For $r$, the reputation bonus is defined as $\mu\mathrm{max}\left(e_{i}-\overline{e},0\right)$.
By combining all the incentives together, we obtain each player's final payoff function for the open review case:
\begin{multline}
\forall \left(e_{i},t_{i}\right)\in E\times T,\\ \pi^{*}_{a}\left(e_{i},t_{i}\right)=\begin{cases}1+\alpha\left(1-e_{i}\right)+\epsilon e_{i}-\beta e_{i}+\mu\mathrm{max}\left(t_{i}-\overline{e},0\right)\mbox{ if }e_{i}\geq t_{i};\\
0-\beta e_{i}\mbox{ otherwise}.
\end{cases}
\end{multline}
\begin{multline}
 \forall \left(e_{i},t_{i}\right)\in E\times T,\\
 \pi_{r}^{*}\left(e_{i},t_{i}\right)=\begin{cases}1+\delta\left(e_{i}-t_{i}\right)+\mu\mathrm{max}\left(e_{i}-\overline{e},0\right)\mbox{ if }e_{i}\geq t_{i};\\
 1\mbox{ otherwise}.
 \end{cases}
\end{multline}

Formally, we thus have an almost identical situation to the double-blind review case. The only difference is that we incorporate the effects of greater transparency on the utilities of the authors and reviewers via the $\mu$ terms. Our principal topic of interest, i.e. the differences of incentives under double-blind and open review in our model, are thus largely reduced to whether values of $\mu$ greater than zero make a substantive difference in the evolutionary expansion of this base game in our full agent-based model.

\subsection{The Agent-based model}

There are two main reasons why we adopt a simulation approach rather than an analytical approach. Firstly, we want to model boundedly rational players, to reflect the unquantified uncertainties and general difficulty of making decisions in a complex and opaque environment like the publication market. Consequently, we shall adopt a probabilistic model where there is a stochastic element in authors' selections of publication strategies. It is hard to determine the ultimate evolution of such a sequence of games using analytical tools, so we use simulations. Secondly, we are principally interested in systematising the incentives of authors and reviewers, as well as their dynamic tendencies. We also hope to stimulate more work in this area by providing a basis for further research that could be turned into more realistic and useful models whose analytical properties might be pertinent for policy choices. Our model is merely an exploratory tool towards these objectives, and therefore its precise formal properties are tangential to our goals.

From an author's perspective, reviewers are randomly assigned to them by editors. In choosing a research strategy (more or less effort put into a particular paper) authors must consider a wide range of information, including both their own experiences and those of their peers. To incorporate this situation into our model, we use an agent-based model in which authors can update their strategies. The model represents the interaction between a population of authors and a population of reviewers who are repeatedly matched to play the author-reviewer game. Each author is initially assigned one of the six effort levels from the set $E=\left(0.3, 0.4, 0.5, 0.6, 0.7, 0.8\right)$. Each reviewer is initially assigned one of the six acceptance threshold from the set $T=\left(0.2, 0.3, 0.4, 0.5, 0.6, 0.7\right)$. This setup ensures that each author is accepted by at least two types of reviewers, to reflect the fact that authors are uncertain about what reviewers will require. We set both populations to contain 1800 agents: one advantage of this choice is that each author has a wide pool of peers who might adopt very different strategies. Across most disciplines, the number of contributing authors has grown from small community to a larger population in the last 60 years \citep{Hamermesh2013}.

In the initial population state, each strategy and effort level is assigned to 300 agents. This is done to eliminate any artificial bias which may act in favour of certain types of agents. We use a probabilistic strategy revision procedure: in every round of the game, the probability of each agent getting assigned a strategy revision opportunity is $0.122$. This number means that, on average, around 219 agents have the opportunity to revise strategies in each round\footnote{We used submissions to economics journals as an inspiration for this figure. While data for the total number of submissions across all economics journals is unavailable, the top five journals received 1160 submissions each in 2011. The number of submissions for low quality journals would naturally be far less \citep{Card2013}.}. The strategy revision procedure is imitative. Once the agent has an opportunity to revise the strategy, he/she compiles a list of $n$ candidates which includes $n-1$ randomly picked agents and the strategy revising agent himself. Thus, strategies which are used by a larger number of agents are more likely to be used by the randomly chosen candidates than the less popular strategies. This reflects the power of following the crowd: one reason why authors might do so is that the popularity of a strategy might be due to its past evolutionary success.

Each candidate plays the agent-reviewer game against every type of opponent, and the strategy revising agent observes the average payoff obtained by each candidate. The strategy revising agent thus obtains a candidate record $\left(s,\overline{\pi}\right)$ \textemdash{} a $n\times 2$ matrix where the first column $s:=\left\{s_{h}\right\}^{n}_{h=1}$ is the strategy record, the second column $\overline{\pi}:=\left\{\overline{\pi}_{h}\right\}^{n}_{h=1}$ is the average payoff record, and the $h$th row of the record is a pair $\left(s_{h},\overline{\pi}_{h}\right)$ that describes the strategy played and the average payoff obtained at revising agent's $h$th observation. If strategy revising agent chooses from finite set of strategies $S=\left(1,...,m\right)$ and uses strategy $i\in S$, the logit choice protocol\footnote{A logit choice model adds an element of randomness to a best response learning rule. To act in accordance with the latter type of rule, agents assign a probability to a strategy based on its past relative performance, and the rule tells them how to make a decision given that strategies' probabilities. A logit choice model adds a noise factor, $\eta$, that an agent will choose a random strategy. Put another way, $\eta$ is a noise term that, in our model, proxies the bounded rationality in agents' strategic response to new data over time. For extensive discussion, see \citep{Blume1997,fudenberg2007,Hofbauer2007}.} defines the probability of agent switching to strategy $j\in S$ as  
\begin{equation}
\sigma_{ij}\left(s,\overline{\pi}\right)=\frac{\sum_{h:s_{h}=j}\mathrm{exp}\left(\eta^{-1}\overline{\pi}_{h}\right)}{\sum_{k\in S}\sum_{h:s_{h}=k}\mathrm{exp}\left(\eta^{-1}\overline{\pi}_{h}\right)}.    
\end{equation}

For candidate selection, we assume that $n=31$. This is an approximately realistic figure: researchers can assess the effort levels of their peers, but not too many of them. For example, by reading an article in depth, one can obtain a sense of how much effort the authors put into the article, but one cannot do this for a large number of articles in a given time-span. We set a low logit noise level $\eta=0.044$. In addition, we assume that each player may switch to any strategy randomly, and the probability of this event is set to $0.008$. The low logit noise level and low probability of random strategy change represent our assumption that players are intelligent and deliberate, so that they rarely make mistakes when choosing their effort level. Each simulation was run for 13000 rounds. This large number of rounds helps to identify the long-run tendencies of a particular game. For the simulations, we assumed some of the parameter values in the author-reviewer game to be fixed. The values of $\alpha$ and $\beta$ were set to $0.1$. We used four values for parameter $\epsilon$ \textemdash{} 0.1, 0.2, 0.3, and 0.4. For each value of $\epsilon$, we used four values of parameter $\delta$ \textemdash{} 0, 0.1, 0.2 and 0.3, and five values of parameter $\mu$ \textemdash{} 0, 0.2, 0.4, 0.8 and 1.6. Therefore, by keeping costs fixed, we can isolate the impact of the institutional environment and how it shapes reputation effects. For simulations, we used Abed simulation package for NetLogo developed by \citep{Izquierdo2019}. 

\subsection{Simulation results}
We now present results obtained through the simulations that we scheduled on the NetLogo software.

Prior to the simulations, for each value of $\epsilon$ and according to all the combinations of parameters we considered (namely $\delta$ and $\mu$), we computed 20 payoff matrices - all bearing dimensions $6\times6$ - by a Python 3 script (see, for instance, two examples respectively in \textbf{table (\ref{table:payoff_matr_eps_0.2})} and \textbf{table (\ref{table:payoff_matr_eps_0.3})}). Using the Python library Nashpy \citep{Knight2018}, we investigated the existence and number of Nash equilibria in pure strategies in every payoff matrix. For those games computed for $\epsilon=\{0.1,0.2\}$ we always found multiple Nash equilibria in pure strategies, whereas for the others calculated for $\epsilon=\{0.3,0.4\}$ we noted a different structure: except all cases marked by $\delta=0$, which had 6 Nash equilibria, all other matrices were characterized by having a unique Nash equilibrium in pure strategies.

\begin{figure} [!htb]
  \caption{\textbf{A strategy distribution for authors}. Here we give an example of a strategies distribution among authors. In panel \textbf{(\ref{fig:1A})} strategies are graphically represented with green color marked by number 1 being the highest effort level and red color marked by number 6 being the lowest one, whereas in panel \textbf{(\ref{fig:1B})} a table based on the same data is shown}
  \centering
  \begin{subfigure}[t]{\textwidth}
  \centering
  \includegraphics[width=0.80\linewidth]{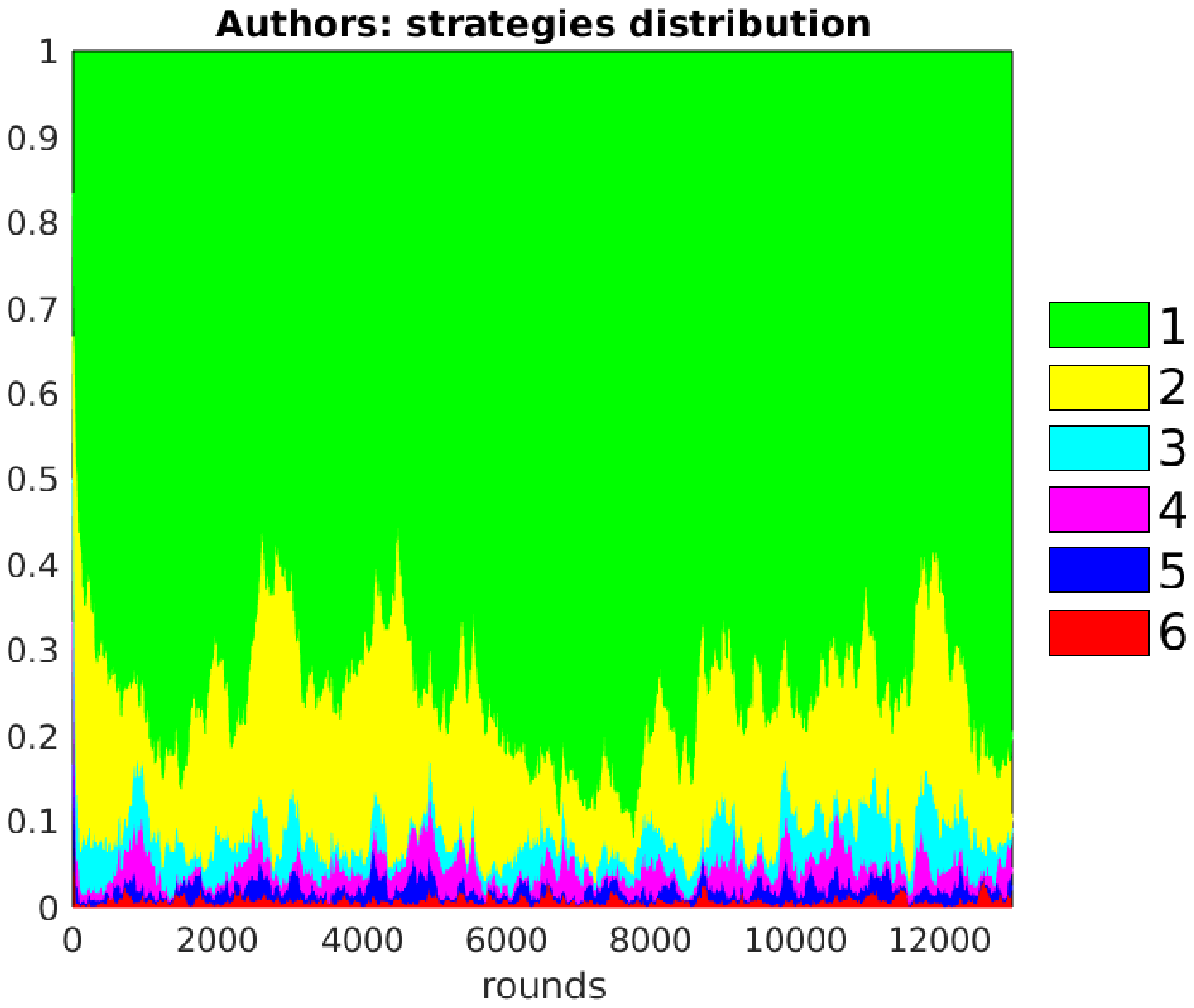}
  \caption{Graph for an Open Review system for $\epsilon=0.3$ with $(\delta,\mu)=(0.3,1.6)$ ruled by matrix \textbf{(b)} shown in \textbf{table (\ref{table:payoff_matr_eps_0.3})}. In this picture, strategies percentages (Y-axis) are plotted against rounds (X-axis)}
  \hfill
  \label{fig:1A}       % Give a unique label
  \end{subfigure}

  \begin{subfigure}[t]{1\textwidth}
  \resizebox{1\textwidth}{!}{%
  \begin{tabular}{@{}l|l|l|l|l|l|l|l|l|@{}}
  \cmidrule(l){2-7}
  & \multicolumn{6}{l|}{Strategies} \\ \midrule
  \multicolumn{1}{|l|}{Data}&1& 2 & 3 & 4 & 5 & 6 & $\overline{S}$ & $\mathbb{E}[E]$\\ \midrule
  \multicolumn{1}{|l|}{Last 3000 rounds} & 71.74 $\pm$ 6.72 & 17.26 $\pm$ 5.31 & 5.31 $\pm$ 2.37 & 2.85 $\pm$ 1.63 & 1.73 $\pm$ 1.09 & 1.12 $\pm$ 0.71 & 1.490 & 0.751 \\
  \midrule
  \multicolumn{1}{|l|}{Entire history} & 73.61 $\pm$ 7.84 & 17.15 $\pm$ 6.72 & 3.93 $\pm$ 2.19 & 2.65 $\pm$ 1.72 & 1.62 $\pm$ 1.18 & 1.05 $\pm$ 0.76 & 1.447 & 0.755 \\
  \bottomrule
  \end{tabular}%
  }
  \vspace{0.25cm}
  \caption{Table presenting mean values and standard deviations (in percentage terms) computed for each strategy over time (rounds) as shown in panel \textbf{(\ref{fig:1A})}. Last two columns report a weighted average for strategies and the expected value of authors effort levels. Both measures are presented for the full set of data and for the last 3000 rounds}
  \hfill
  \label{fig:1B}       % Give a unique label
  \end{subfigure}
  \label{fig:1}
\end{figure}

After the Nash equilibria check, for each payoff matrix we ran simulations for 13000 rounds, and we assessed a discrete probability distribution $P$ for the 6 strategies we considered. For each simulation, we performed several analyses. For instance, see \textbf{figure (\ref{fig:1})}, where we considered an Open Review system from the authors' viewpoint. In all simulations, the final goal was to compute the expected value of authors effort levels as
\begin{equation}
    \mathbb{E}[E]=\sum_{i=1}^6e_ip_i
\end{equation}
and, similarly, the related expected value of reviewers' threshold levels. For every simulation, we considered the entire history and the last 3000 rounds, amounting at about the 23\% of the full data-set. We report the expected values for $\epsilon=\{0.1,0.2,0.3,0.4\}$ for authors in \textbf{tables (\ref{table:expected_values_eps_0.1_authors_all},\ref{table:expected_values_eps_0.2_authors_all},\ref{table:expected_values_eps_0.3_authors_all},\ref{table:expected_values_eps_0.4_authors_all})} for the entire data-set and \textbf{tables (\ref{table:expected_values_eps_0.1_authors_last_3000},\ref{table:expected_values_eps_0.2_authors_last_3000},\ref{table:expected_values_eps_0.3_authors_last_3000},\ref{table:expected_values_eps_0.4_authors_last_3000})} with respect to the last 3000 rounds, and similarly for reviewers in \textbf{tables (\ref{table:expected_values_eps_0.1_reviewers_all},\ref{table:expected_values_eps_0.2_reviewers_all},\ref{table:expected_values_eps_0.3_reviewers_all},\ref{table:expected_values_eps_0.4_reviewers_all})} and \textbf{tables (\ref{table:expected_values_eps_0.1_reviewers_last_3000},\ref{table:expected_values_eps_0.2_reviewers_last_3000},\ref{table:expected_values_eps_0.3_reviewers_last_3000},\ref{table:expected_values_eps_0.4_reviewers_last_3000})}.

\begin{figure}[!htb]
    \caption{\textbf{Strategy distributions for authors}. On left panels, figures representing interactions under Double Blind system are shown, whereas on right panels examples stemming from Open Review mechanism are reported. Strategies percentages (Y-axis) are pictured against rounds (X-axis). At the same time, in the upper panels (\subref{fig:pop_1_fig_1}) and (\subref{fig:pop_1_fig_2}) we display authors' strategy distributions for $\epsilon=0.1$ and in lower panels (\subref{fig:pop_1_fig_3}) and (\subref{fig:pop_1_fig_4}) for $\epsilon=0.4$}
    \label{fig:2}
    \centering
    \begin{subfigure}[t]{0.45\textwidth}
        \centering
        \includegraphics[width=1.15\linewidth]{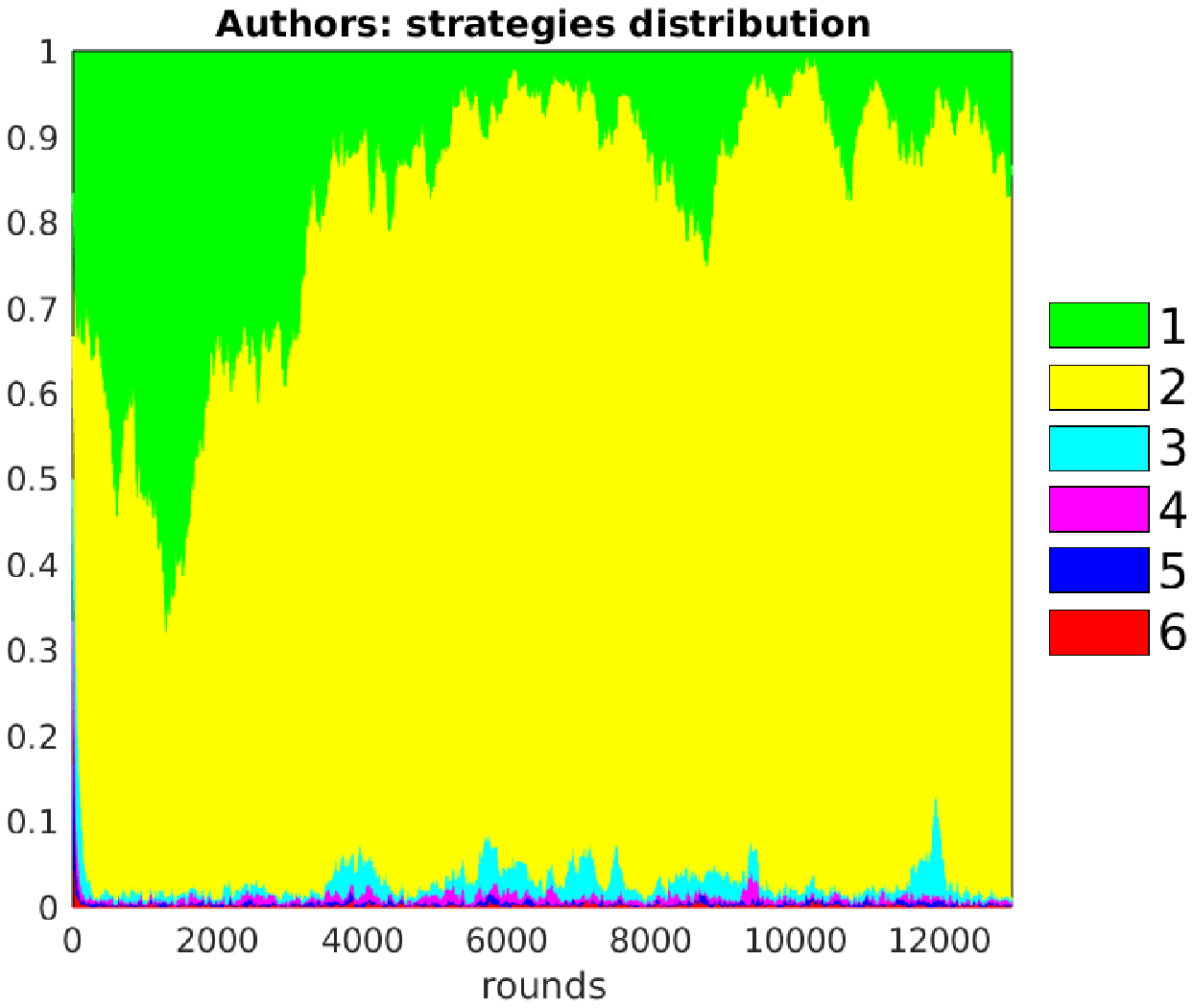} 
        \caption{$(\delta,\mu)=(0,0)$} \label{fig:pop_1_fig_1}
    \end{subfigure}
    \hfill
    \begin{subfigure}[t]{0.45\textwidth}
        \centering
        \includegraphics[width=1.15\linewidth]{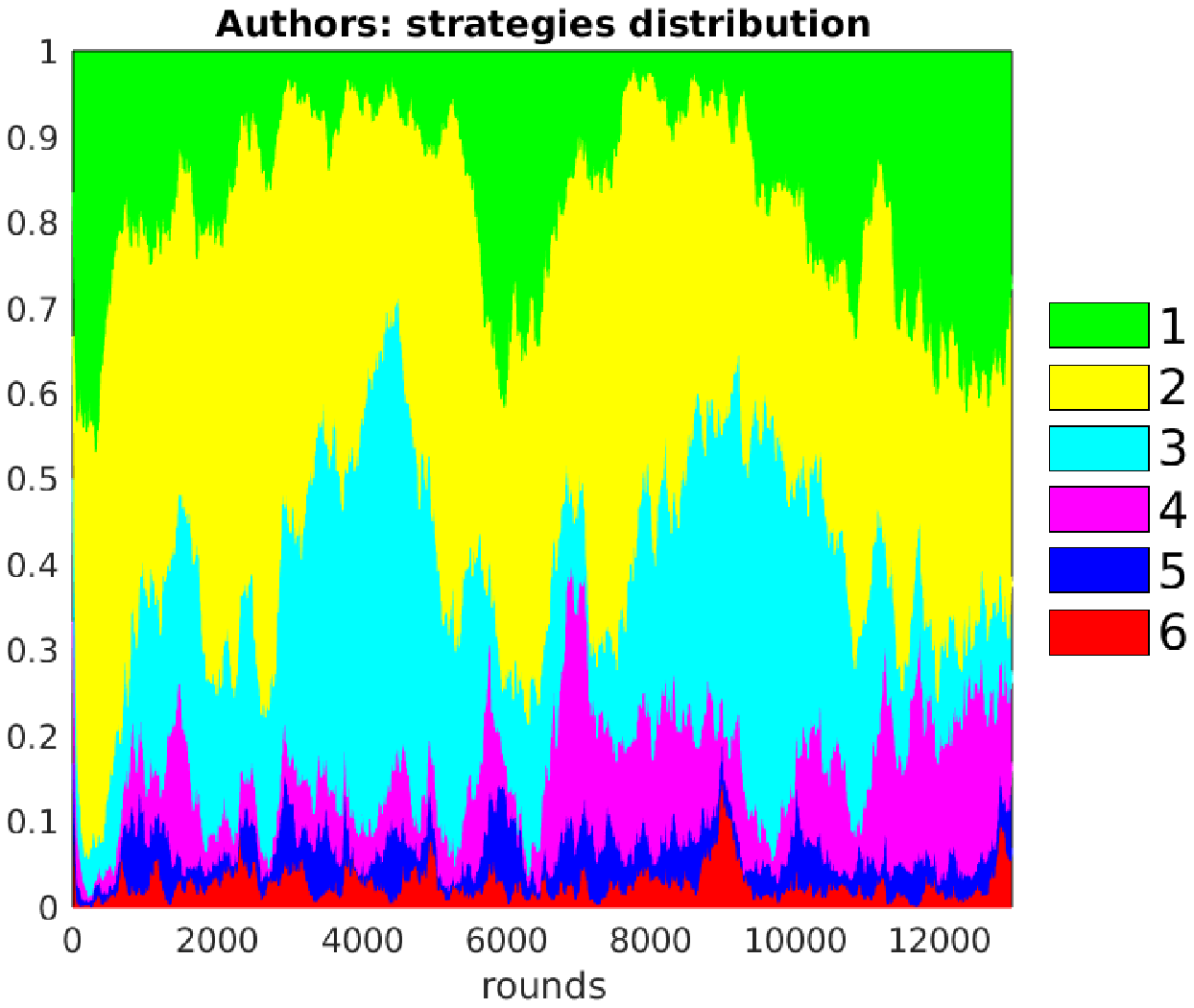} 
        \caption{$(\delta,\mu)=(0.3,1.6)$} \label{fig:pop_1_fig_2}
    \end{subfigure}

    \vspace{1cm}
    \begin{subfigure}[t]{0.45\textwidth}
        \centering
        \includegraphics[width=1.15\linewidth]{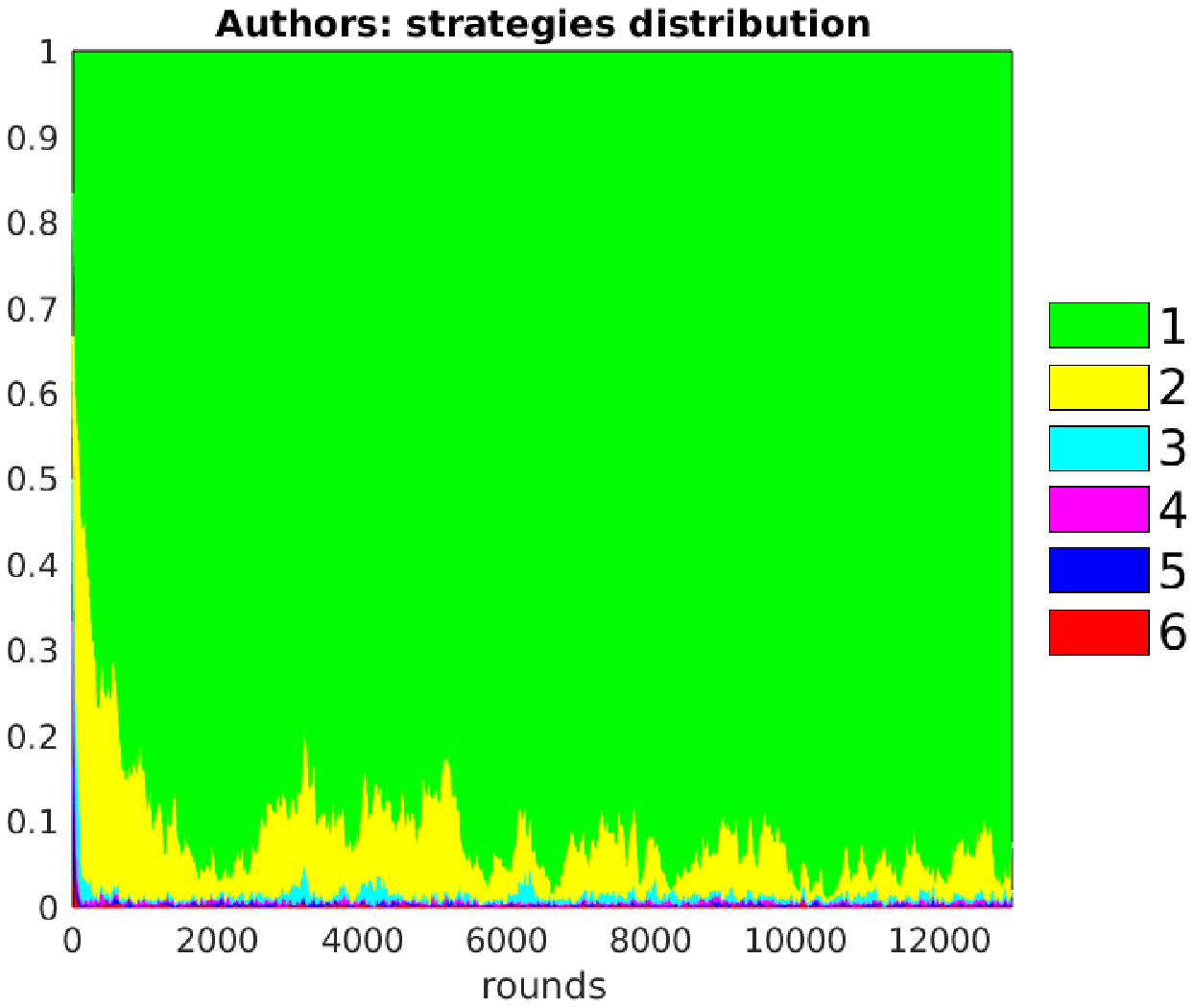} 
        \caption{$(\delta,\mu)=(0,0)$} \label{fig:pop_1_fig_3}
    \end{subfigure}
    \hfill
    \begin{subfigure}[t]{0.45\textwidth}
        \centering
        \includegraphics[width=1.15\linewidth]{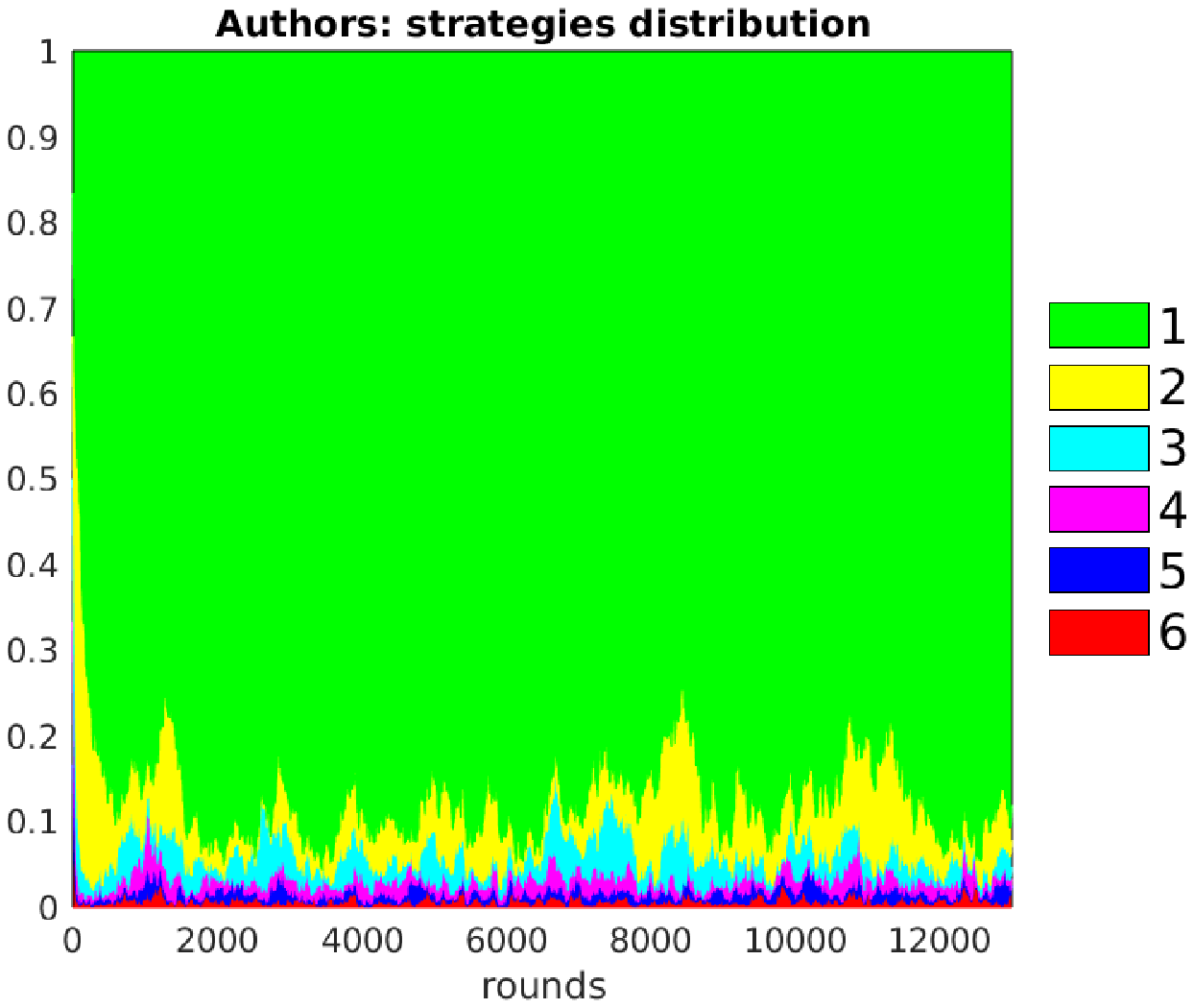} 
        \caption{$(\delta,\mu)=(0.3,1.6)$} \label{fig:pop_1_fig_4}
    \end{subfigure}
\end{figure}

\begin{figure}[!htb]
    \caption{\textbf{Strategy distributions for reviewers}. Left and right panels are organized mirroring for reviewers what is presented for authors in \textbf{figure (\ref{fig:2})}. Similarly, in upper panels (\subref{fig:pop_2_fig_1}) and (\subref{fig:pop_2_fig_2}) we display reviewers' strategy distributions for $\epsilon=0.1$ and in lower panels (\subref{fig:pop_2_fig_3}) and (\subref{fig:pop_2_fig_4}) for $\epsilon=0.4$}
    \label{fig:3}
    \centering
    \begin{subfigure}[t]{0.45\textwidth}
        \centering
        \includegraphics[width=1.15\linewidth]{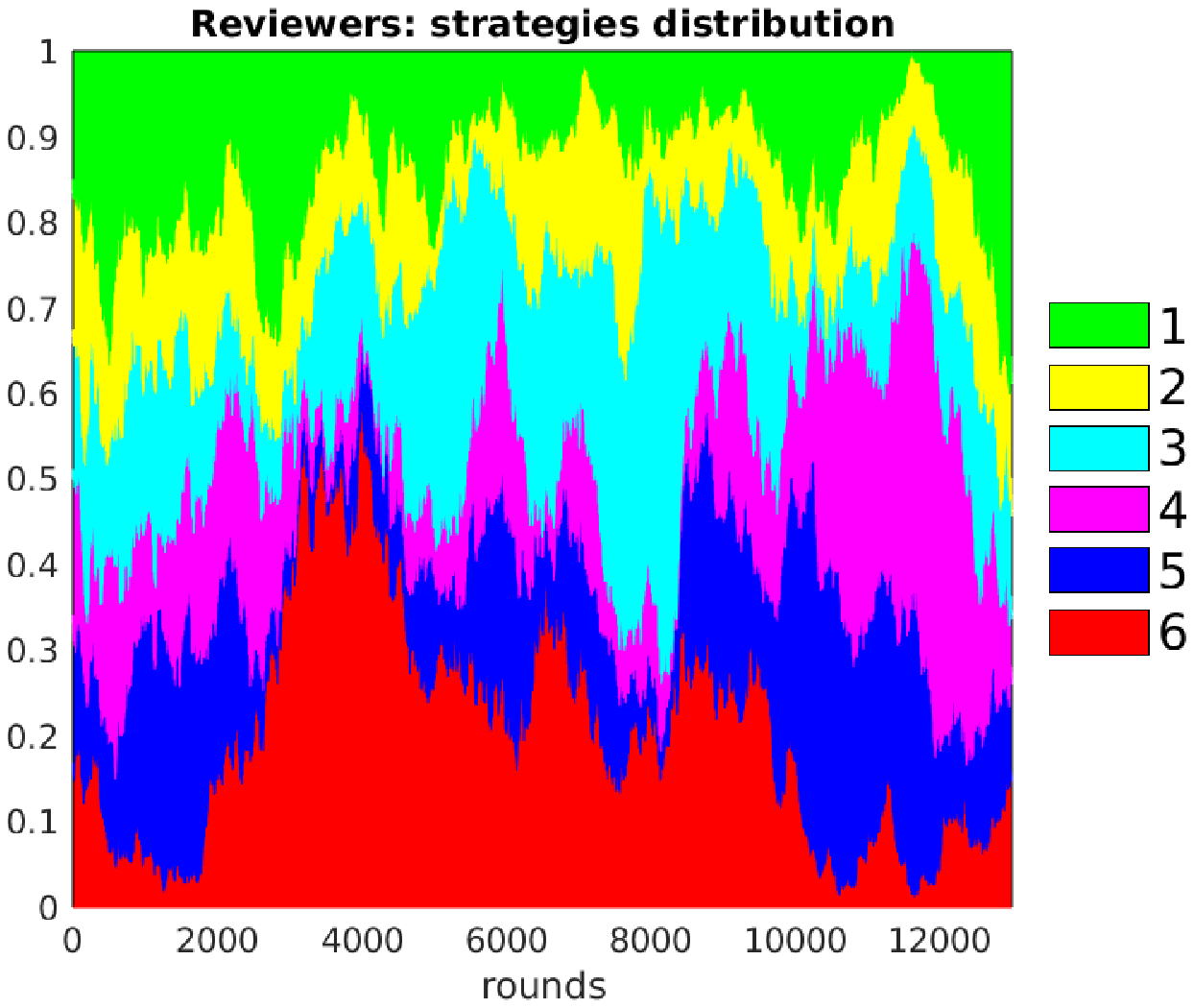} 
        \caption{$(\delta,\mu)=(0,0)$} \label{fig:pop_2_fig_1}
    \end{subfigure}
    \hfill
    \begin{subfigure}[t]{0.45\textwidth}
        \centering
        \includegraphics[width=1.15\linewidth]{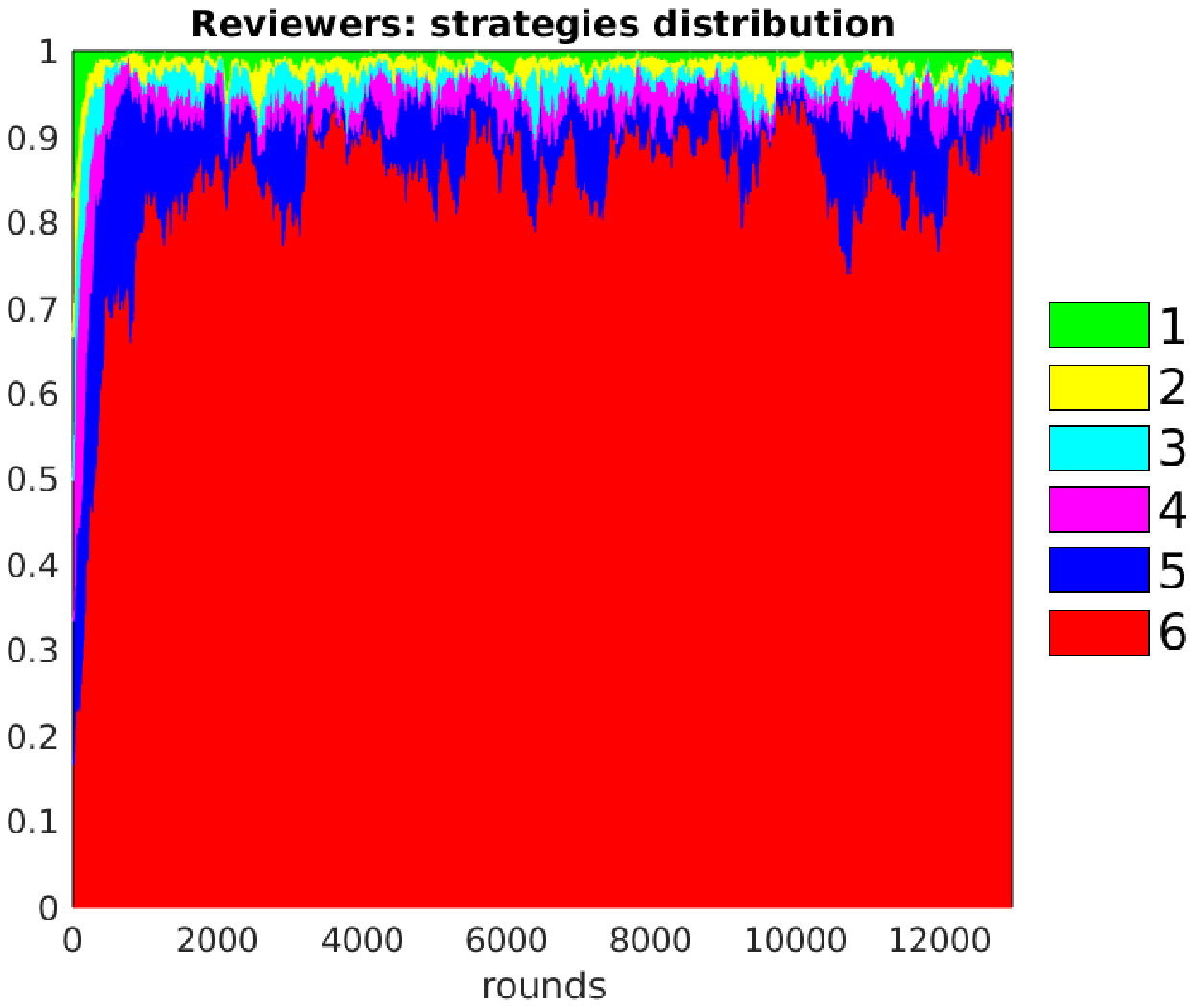} 
        \caption{$(\delta,\mu)=(0.3,1.6)$} \label{fig:pop_2_fig_2}
    \end{subfigure}

    \vspace{1cm}
    \begin{subfigure}[t]{0.45\textwidth}
        \centering
        \includegraphics[width=1.15\linewidth]{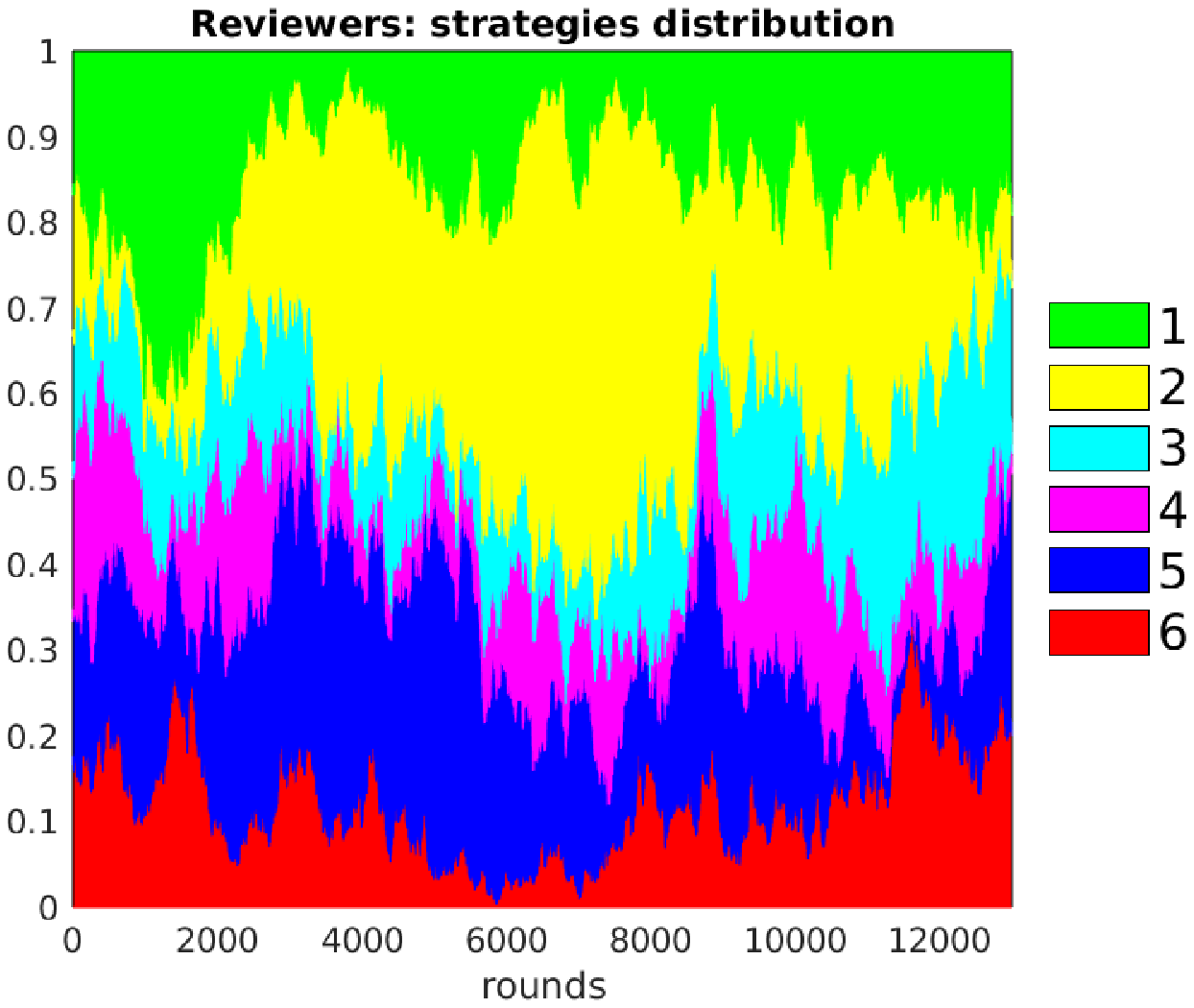} 
        \caption{$(\delta,\mu)=(0,0)$} \label{fig:pop_2_fig_3}
    \end{subfigure}
    \hfill
    \begin{subfigure}[t]{0.45\textwidth}
        \centering
        \includegraphics[width=1.15\linewidth]{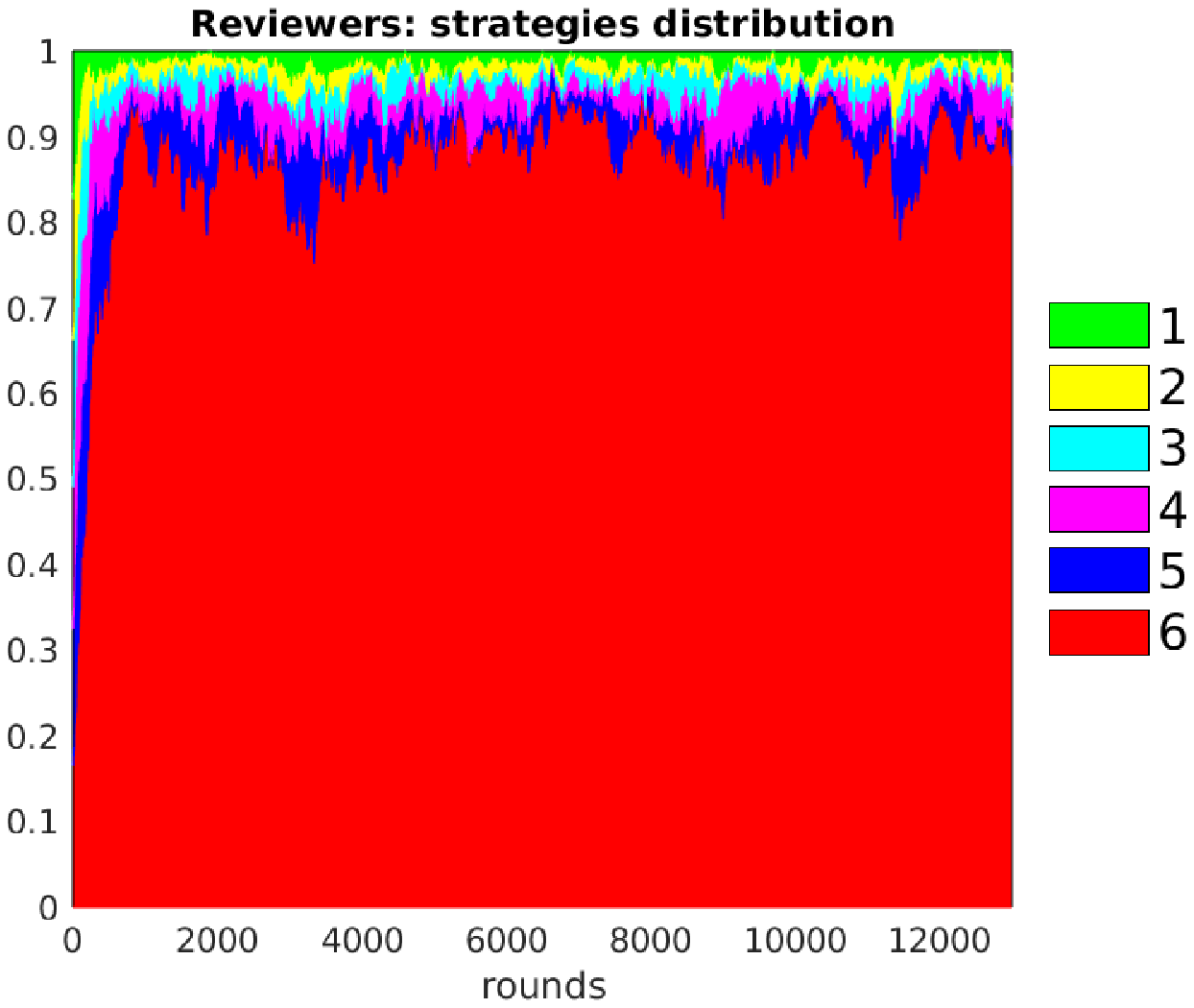} 
        \caption{$(\delta,\mu)=(0.3,1.6)$} \label{fig:pop_2_fig_4}
    \end{subfigure}
\end{figure}

A graphical comparison for authors and reviewers between Double Blind system and Open Review mechanism for different values of $\epsilon$ is shown in \textbf{figure (\ref{fig:2})} and \textbf{(\ref{fig:3})}.

\section{Discussion}
Our model is not intended to be a realistic representation of how peer review actually operates. What it contributes is a novel application of game-theoretic modelling in an area of debate where the discussion of incentives has been mostly informal and where empirical research is difficult.

We turn now to our results. Neither open review nor double-blinding shows a consistent advantage for the purpose of encouraging researchers' effort. There are some cases where one system does significantly better (e.g. in Table 16, when $\delta>0$ and $\mu=0.2$, Open Review very strongly incentivises reviewer effort) but these do not manifest any obvious regularities. Furthermore, we should be wary about interpreting small differences in results, which could easily be the result of particular choices parameters like the noise function $\eta$. If this feature of our model was reflected in the real world, then it could be interpreted as slightly in favour of open review, because there are non-trivial costs to enforcing double-blinding (assuming that such enforcement is possible) with no particular advantages with respect to incentivising research effort\footnote{There might be other advantages, e.g. improving fairness by preventing racial and gender biases.}. Note that this does not mean that \emph{peer review} is ineffective in our model: all papers are peer reviewed under every scenario we examine, so we cannot make any comparisons between peer reviewed science and non-peer reviewed science in our model.

The value of $\delta$ has a fairly consistent (negative) impact on effort levels, although its impact is not always very strong. In particular, the effort levels of both authors and reviewers are almost always greater when $\delta$ = 0. Under double-blinding, i.e. $\mu$ = 0, a higher value of $\delta$ always leads to lower effort, and this impact is especially strong (though very non-linear) for reviewers. This result is easy to interpret: when reviewers' effort is completely free, then they will provide it in abundance; when they do not, they will economise on it. However, since authors' behaviour is not a simple reflection of reviewers' effort levels, it can still be advantageous for authors to put effort into their papers even if reviewers do not put effort into their reviews. High values of $\epsilon$ and/or $\mu$ can offset the deleterious consequences of low values of $\delta$ by providing robust incentives in favour of high effort for authors For this reason, there is no neat relationship between $\delta$ and authors' effort levels. Thus, if we imagine that $\delta$ was replaced by a more complex parameter $\chi$ consisting of both (1) reviewer costs and (2) some exogenous compensation for higher effort levels, even reducing $\chi$ to approximately zero would not have a very strong impact on authors' effort levels.

Surprisingly, an increase in $\mu$ has no straightforward relation to an increase in researcher effort. In particular, under some circumstances, when $\mu$ is large, this strong reputation bonus for high effort reviewing can lead reviewers to become complacent due to general high levels of effort. Thus, a strategy of low-effort papers can have long-run benefits, which reduces long-run effort. In other words, in our model, strong reputation bonuses for high reviewer effort can create an environment that counter-intuitively benefits a strategy of low author effort. Smaller reviewer reputation bonuses can actually increase overall average author effort levels. While this result is perhaps counter-intuitive, it does not appear to be the direct result of some unrealistic assumption in our model, and therefore it is a possibility (which has not been discussed in the existing literature on peer review) that we should consider. If it was reflected in the real world, it would mean that (under open review) there were benefits to not having vast reputation bonuses for higher effort reviewing, because such an environment keeps reviewers ``on their toes'' in their watchdog role against low effort research. It would also imply that the benefits of open review for encouraging high effort research would be, at best, non-robust.

In our model, there are strong structural incentives in favour of low-effort research, in that the rationality of this strategy for authors is insensitive to the peer review system. This seems to reflect a feature of real-world publication markets: over a large number of submissions, even a low-effort researcher has a non-zero probability of ``getting lucky'' with their reviewers, and the utility of obtaining the desired result with low effort can overwhelm the lower probability of acceptance\footnote{Analogously, consider spam email scammers. Since the costs of sending out the emails is so low and the benefits from success are so high, it usually makes sense to send out generic, low-effort scam emails to many people rather than a few carefully crafted and customised attempts.}. Good scientific reviewing, meanwhile, is a positive externality: a reviewer might have a utility bonus from rejecting a bad paper or accepting a good paper, but this will not reflect the full social benefits to the scientific community. Our model cannot tell us whether, in the real world, peer review institutions do not alter these underlying incentives, but they do demonstrate the \emph{possibility} that they do not.

While the peer review system does not make a noticeable difference in our model, $\epsilon$ (authors' reputation bonus) is a reliably important parameter. If $\epsilon$ is higher, then the effort levels for both authors and reviewers rise for any particular values of the other parameters. An example can be seen in \textbf{Figure 2}. Under double-blind review, i.e. when $\delta$ and $\mu$ are zero, then a shift of $\epsilon$ from 0.1 to 0.4 makes a large (though not overwhelming) difference to author effort, as shown in graphs \textbf{(a)} and \textbf{(c)}. This difference between $\epsilon$ = 0.1 and $\epsilon$ = 0.4 is even greater under the particular open review environment that has high values for $\delta$ and $\mu$, as one can see in graphs \textbf{(b)} and \textbf{(d)}. For the full details, see Tables 3 to 18.

This regularity is partly attributable to a structural feature of the model, which we noted earlier in this article in Subsection 2.3. When $\epsilon$ = 0.1 or $\epsilon$ = 0.2, then the game always has multiple Nash equilibria for all the settings of $\delta$ and $\mu$ that we examined. In contrast, when $\epsilon$ = 0.3 or $\epsilon$ = 0.4, the game has a single Nash equilibrium when $\delta$ is non-zero. (It still has multiple Nash equilibria - six - when $\delta$ is zero.) Informally, when $\epsilon$ becomes very low, both authors and reviewers become apathetic (they are indifferent among many outcomes) and therefore the structural incentives for author effort are weaker. It is unsurprising that $\epsilon$ would reliable make a difference, but it is perhaps surprising that it is the only parameter we examine under different settings (i.e. other than parameters like $\alpha$ and $\beta$) that reliably makes a reliable difference.

On an optimistic note, our study identifies a possible strategy that can lead to replicability problems: sending out many low-effort papers, with a good chance of accumulating a superficially impressive research record. Even if this strategy is difficult to disincentivise via changing from double-blinding to open review or vice versa, there are other conceivable ways to discourage it. For instance, consider hiring, tenure, and funding decisions. If authorities making these decisions act in a way to reward high-effort research, \emph{even if it results in a comparatively terse publication record}, then that could have a large impact. This amounts would be analogous to raising $\epsilon$ in our model, i.e. increasing the reputation bonus for author effort. An example of an institutional move in this direct is a feature of the UK government's Research Excellence Framework (previously ``Research Assessment Exercise'') which is a regular nationwide assessment of departments with respect to their suitability for receiving future research funds\footnote{\href{https://www.ref.ac.uk/publications/guidance-on-submissions-201901/}{https://www.ref.ac.uk/publications/guidance-on-submissions-201901/}}. Individual researchers are limited to (co)authorship of a maximum of five research items. Therefore, over the six year period since the last assessment, in 2014, a researcher who published about one high quality research item per year would be more valuable to their department than a researcher who annually published far more low quality research items. In short, peer review is not the only way that low quality research might be discouraged.

One of main purposes of our model is to inspire further research. A natural area for further investigation are the incentives facing editors, whom we did not include in our model. While editors take reviewers' recommendations seriously, their decisions to accept or reject are not purely determined by what reviewers say. Furthermore, editors make decisions about desk-rejections, e.g. whether to reject any empirical paper that was not preregistered. A very long-term ambition would be a general model of the publication market and its effects on the quality of research, but this is very far from a possibility right now.

A limitation of our model is that we treat authors and reviewers as separate populations. In the real world, a scientist will generally act in both roles on many occasions. One way that our model could be more sophisticated would be to address this limitation. A hypothesis worth investigating would be whether the size of the overall population of authors/reviewers affects the degree of interrelation between scientists' author-behaviour and their reviewer-behaviour. For instance, in a small field, a scientist might raise the average effort level by working harder as a reviewer, which may be against their self-interest. In contrast, in a large field, one individual's decisions seems unlikely to affect the field's overall standards. While this hypothesis is plausible, some of our surprising results from the model's current version indicate that we should be sceptical of such intuitions until they are verified by the general tendencies in simulations or by analytical results.

\section{Conclusions}

We have not sought to advance simulation methods at a technical level. Yet this familiarity is a strength, because our topic is so new that it would be risky to combine both novel methods and a novel domain; our use of well-grounded tools removes this risk. Methodologically, our study has the inevitable shortcomings of simulation research, in contrast to analytical proofs. However, since we are exploring new territory, it is probably best to leave such proofs until more realistic and well-calibrated models have been developed. At this point, simulations can provide surprising information about how some of the incentives in peer review can interact with each other.

Our results illustrate an important general point about complex institutional contexts and their reform: even in simplified models, it is hard to predict how incentives will interact or how they will react to regime change. The high fallibility of human intuition in these areas means that game-theoretic modelling has much to add. While game-theoretic results are inevitably just possibility results (agents might not behave strategically) these can serve a powerful corrective function to our intuitions about how scientists will react to institutional reforms, especially the heated intuitions we often have in the replication crisis debates.

Over many years, Francesc Carreras Escobar has enriched game theory, both in its fundamentals and its applications. His use of game theory to understand institutions such as European political bodies \citep{carreras2016} and the Catalonian parliament \citep{carreras1988} is an inspiring indication of the power of game theory. While we have no pretence of having achieved such applicable models in this paper, we hope to have laid the foundations, in the game-theoretic study of peer review, for matching his successes.
\vspace{0.5cm}
\\
\textbf{Acknowledgements}\\
We thank Todd Stambaugh (City University of New York, USA) for his helpful inputs into the paper, especially the choice of topic.
\\
\\
\textbf{Funding}\\
Mantas Radzvilas, Francesco De Pretis, William Peden, and Barbara Osimani acknowledge funding from the European Research Council (GA n. 639276) and the Marche Polytechnic University (Ancona, Italy).
\\
\\
\textbf{Conflicts of interest}\\
Authors declare no conflicts of interest.
\\
\\
\textbf{Availability of data and material (data transparency)}\\
Authors are available to provide employed data for this manuscript at their request.
\\
\\
\textbf{Code availability (software application or custom code)}\\
Authors are available to provide employed code for this manuscript at their request.
\\
\\
\textbf{Authors' Contributions}\\
Mantas Radzvilas developed the model, primarily ran the simulations, and co-wrote Section 2. Francesco De Pretis developed the simulation process, ran simulations, co-wrote Section 2, and developed the graphics. William Peden wrote Sections 1 and 3, and co-wrote Section 2. Daniele Tortoli developed the measurement of the simulation results, developed the graphics, and ran simulations. Barbara Osimani suggested the topic and provided continuous discussions on the article.

%\bibliographystyle{apa}

\bibliographystyle{abbrvnat}

\bibliography{references.bib}   % name your BibTeX data base

%\begin{thebibliography}{}
% etc
%\end{thebibliography}

\hfill
\clearpage

\section*{FIGURES}

\pagebreak
\hfill
\newpage
\pagebreak
\hfill
\newpage

\section*{TABLES}

\begin{table}[!htb]
    \caption{Examples of payoff matrices for $\epsilon=0.2$. Matrix \textbf{(a)} is defined for a Double Blind system and it is characterized by having 26 Nash equilibria in pure strategies, whereas matrix \textbf{(b)} is calculated for an Open Review mechanism and shows 6 Nash equilibria in pure strategies.}
    \label{table:payoff_matr_eps_0.2}
    \begin{subtable}[t]{1\textwidth}
    \setlength{\extrarowheight}{2pt}
    \resizebox{\textwidth}{!}{%
    \begin{tabular}{*{36}{c|}}
      \multicolumn{2}{c}{} & \multicolumn{6}{c}{Reviewer}\\\cline{3-8}
      \multicolumn{1}{c}{} &  & $S1$  & $S2$ & $S3$  & $S4$ & $S5$  & $S6$\\\cline{2-8}
      \multirow{6}*{Author}  & $S1$ & $(1.10,1.00)$ & $(1.10,1.00)$ & $(-0.03,1.00)$ & $(-0.03,1.00)$ & $(-0.03,1.00)$ & $(-0.03,1.00)$\\\cline{2-8}
      & $S2$ & $(1.10,1.00)$ & $(1.10,1.00)$ & $(1.10,1.00)$ & $(-0.04,1.00)$ & $(-0.04,1.00)$ & $(-0.04,1.00)$\\\cline{2-8}
      & $S3$ & $(1.10,1.00)$ & $(1.10,1.00)$ & $(1.10,1.00)$ & $(1.10,1.00)$ & $(-0.05,1.00)$ & $(-0.05,1.00)$\\\cline{2-8}
      & $S4$ & $(1.10,1.00)$ & $(1.10,1.00)$ & $(1.10,1.00)$ & $(1.10,1.00)$ & $(1.10,1.00)$ & $(-0.06,1.00)$\\\cline{2-8}
      & $S5$ & $(1.10,1.00)$ & $(1.10,1.00)$ & $(1.10,1.00)$ & $(1.10,1.00)$ & $(1.10,1.00)$ & $(1.10,1.00)$\\\cline{2-8}
      & $S6$ & $(1.10,1.00)$ & $(1.10,1.00)$ & $(1.10,1.00)$ & $(1.10,1.00)$ & $(1.10,1.00)$ & $(1.10,1.00)$\\\cline{2-8}
      \end{tabular}%
      }
      \hfill
      \label{table:payoff_matr_eps_0.2_a}
      \caption{Payoff matrix computed for $(\delta,\mu)=(0,0)$}
      \end{subtable}
      
    \begin{subtable}[t]{1\textwidth}
    \setlength{\extrarowheight}{2pt}
    \resizebox{\textwidth}{!}{%
    \begin{tabular}{*{36}{c|}}
      \multicolumn{2}{c}{} & \multicolumn{6}{c}{Reviewer}\\\cline{3-8}
      \multicolumn{1}{c}{} &  & $S1$  & $S2$ & $S3$  & $S4$ & $S5$  & $S6$\\\cline{2-8}
      \multirow{6}*{Author}  & $S1$ & $(1.10,1.03)$ & $(1.10,1.00)$ & $(-0.03,1.00)$ & $(-0.03,1.00)$ & $(-0.03,1.00)$ & $(-0.03,1.00)$\\\cline{2-8}
      & $S2$ & $(1.10,1.06)$ & $(1.10,1.03)$ & $(1.10,1.00)$ & $(-0.04,1.00)$ & $(-0.04,1.00)$ & $(-0.04,1.00)$\\\cline{2-8}
      & $S3$ & $(1.10,1.17)$ & $(1.10,1.14)$ & $(1.10,1.11)$ & $(1.18,1.08)$ & $(-0.05,1.00)$ & $(-0.05,1.00)$\\\cline{2-8}
      & $S4$ & $(1.10,1.36)$ & $(1.10,1.33))$ & $(1.10,1.30))$ & $(1.18,1.27))$ & $(1.34,1.24)$ & $(-0.06,1.00)$\\\cline{2-8}
      & $S5$ & $(1.10,1.55)$ & $(1.10,1.52)$ & $(1.10,1.49)$ & $(1.18,1.46)$ & $(1.34,1.43)$ & $(1.50,1.40)$\\\cline{2-8}
      & $S6$ & $(1.10,1.74)$ & $(1.10,1.71)$ & $(1.10,1.68)$ & $(1.18,1.65)$ & $(1.34,1.62)$ & $(1.50,1.59)$\\\cline{2-8}
      \end{tabular}%
      }
      \hfill
      \label{table:payoff_matr_eps_0.2_b}
      \caption{Payoff matrix computed for $(\delta,\mu)=(0.3,1.6)$}
      \end{subtable}
  \end{table}
  
\begin{table}[!htb]
    \caption{Examples of payoff matrices for $\epsilon=0.3$. Matrix \textbf{(a)} is defined for a Double Blind system and it is characterized by having 6 Nash equilibria in pure strategies, whereas matrix \textbf{(b)} is calculated for an Open Review mechanism and owns a unique Nash equilibrium in pure strategies.}
    \label{table:payoff_matr_eps_0.3}
    \begin{subtable}[t]{1\textwidth}
    \setlength{\extrarowheight}{2pt}
    \resizebox{\textwidth}{!}{%
    \begin{tabular}{*{36}{c|}}
      \multicolumn{2}{c}{} & \multicolumn{6}{c}{Reviewer}\\\cline{3-8}
      \multicolumn{1}{c}{} &  & $S1$  & $S2$ & $S3$  & $S4$ & $S5$  & $S6$\\\cline{2-8}
      \multirow{6}*{Author}  & $S1$ & $(1.13,1.00)$ & $(1.13,1.00)$ & $(-0.03,1.00)$ & $(-0.03,1.00)$ & $(-0.03,1.00)$ & $(-0.03,1.00)$\\\cline{2-8}
      & $S2$ & $(1.14,1.00)$ & $(1.14,1.00)$ & $(1.14,1.00)$ & $(-0.04,1.00)$ & $(-0.04,1.00)$ & $(-0.04,1.00)$\\\cline{2-8}
      & $S3$ & $(1.15,1.00)$ & $(1.15,1.00)$ & $(1.15,1.00)$ & $(1.15,1.00)$ & $(-0.05,1.00)$ & $(-0.05,1.00)$\\\cline{2-8}
      & $S4$ & $(1.16,1.00)$ & $(1.16,1.00)$ & $(1.16,1.00)$ & $(1.16,1.00)$ & $(1.16,1.00)$ & $(-0.06,1.00)$\\\cline{2-8}
      & $S5$ & $(1.17,1.00)$ & $(1.17,1.00)$ & $(1.17,1.00)$ & $(1.17,1.00)$ & $(1.17,1.00)$ & $(1.17,1.00)$\\\cline{2-8}
      & $S6$ & $(1.18,1.00)$ & $(1.18,1.00)$ & $(1.18,1.00)$ & $(1.18,1.00)$ & $(1.18,1.00)$ & $(1.18,1.00)$\\\cline{2-8}
      \end{tabular}%
      }
      \hfill
      \label{table:payoff_matr_eps_0.3_a}
      \caption{Payoff matrix computed for $(\delta,\mu)=(0,0)$}
      \end{subtable}
      
    \begin{subtable}[t]{1\textwidth}
    \setlength{\extrarowheight}{2pt}
    \resizebox{\textwidth}{!}{%
    \begin{tabular}{*{36}{c|}}
      \multicolumn{2}{c}{} & \multicolumn{6}{c}{Reviewer}\\\cline{3-8}
      \multicolumn{1}{c}{} &  & $S1$  & $S2$ & $S3$  & $S4$ & $S5$  & $S6$\\\cline{2-8}
      \multirow{6}*{Author}  & $S1$ & $(1.13,1.03)$ & $(1.13,1.00)$ & $(-0.03,1.00)$ & $(-0.03,1.00)$ & $(-0.03,1.00)$ & $(-0.03,1.00)$\\\cline{2-8}
      & $S2$ & $(1.14,1.06)$ & $(1.14,1.03)$ & $(1.14,1.00)$ & $(-0.04,1.00)$ & $(-0.04,1.00)$ & $(-0.04,1.00)$\\\cline{2-8}
      & $S3$ & $(1.15,1.17)$ & $(1.15,1.14)$ & $(1.15,1.11)$ & $(1.23,1.08)$ & $(-0.05,1.00)$ & $(-0.05,1.00)$\\\cline{2-8}
      & $S4$ & $(1.16,1.36)$ & $(1.16,1.33))$ & $(1.16,1.30))$ & $(1.24,1.27))$ & $(1.40,1.24)$ & $(-0.06,1.00)$\\\cline{2-8}
      & $S5$ & $(1.17,1.55)$ & $(1.17,1.52)$ & $(1.17,1.49)$ & $(1.25,1.46)$ & $(1.41,1.43)$ & $(1.57,1.40)$\\\cline{2-8}
      & $S6$ & $(1.18,1.74)$ & $(1.18,1.71)$ & $(1.18,1.68)$ & $(1.26,1.65)$ & $(1.42,1.62)$ & $(1.58,1.59)$\\\cline{2-8}
      \end{tabular}%
      }
      \hfill
      \label{table:payoff_matr_eps_0.3_b}
      \caption{Payoff matrix computed for $(\delta,\mu)=(0.3,1.6)$}
      \end{subtable}
  \end{table}

\begin{table}[!htb]
\caption{Expected values of effort levels (authors) computed for $\epsilon=0.1$ over an entire round of simulation (13k rounds).}
\label{table:expected_values_eps_0.1_authors_all}
\centering
\begin{tabular}{@{}l|l|l|l|l|l|@{}}
\cmidrule(l){2-6}
                     & Double Blind & \multicolumn{4}{c|}{Open Review}  \\ \cmidrule(l){1-6} 
\multicolumn{1}{|c|}{\diagbox{$\delta$}{$\mu$}}                  & \multicolumn{1}{c|}{$\mu=0$} & \multicolumn{1}{c|}{$\mu=0.2$} & \multicolumn{1}{c|}{$\mu=0.4$} & \multicolumn{1}{c|}{$\mu=0.8$} & \multicolumn{1}{c|}{$\mu=1.6$} \\ \midrule
\multicolumn{1}{|l|}{$\delta=0$}  & \multicolumn{1}{|c|} {0.71} & \multicolumn{1}{|c|} {0.709} & \multicolumn{1}{|c|} {0.696} & \multicolumn{1}{|c|} {0.664} & \multicolumn{1}{|c|} {0.64}                                \\ \midrule
\multicolumn{1}{|l|}{$\delta=0.1$}  & \multicolumn{1}{|c|} {0.691} & \multicolumn{1}{|c|} {0.716} & \multicolumn{1}{|c|} {0.693} & \multicolumn{1}{|c|} {0.668} & \multicolumn{1}{|c|} {0.644}                                \\ \midrule
\multicolumn{1}{|l|}{$\delta=0.2$}  & \multicolumn{1}{|c|} {0.658} & \multicolumn{1}{|c|} {0.709} & \multicolumn{1}{|c|} {0.696} & \multicolumn{1}{|c|} {0.655} & \multicolumn{1}{|c|} {0.652}                              \\ \midrule
\multicolumn{1}{|l|}{$\delta=0.3$}  & \multicolumn{1}{|c|} {0.629} & \multicolumn{1}{|c|} {0.712} & \multicolumn{1}{|c|} {0.696} & \multicolumn{1}{|c|} {0.667} & \multicolumn{1}{|c|} {0.647}                                \\ \bottomrule
\end{tabular}
\end{table}

\begin{table}[!htb]
\caption{Expected values of threshold levels (reviewers) computed for $\epsilon=0.1$ over an entire round of simulation (13k rounds).}
\label{table:expected_values_eps_0.1_reviewers_all}
\centering
\begin{tabular}{@{}l|l|l|l|l|l|@{}}
\cmidrule(l){2-6}
                     & Double Blind & \multicolumn{4}{c|}{Open Review}  \\ \cmidrule(l){1-6} 
\multicolumn{1}{|c|}{\diagbox{$\delta$}{$\mu$}}                  & \multicolumn{1}{c|}{$\mu=0$} & \multicolumn{1}{c|}{$\mu=0.2$} & \multicolumn{1}{c|}{$\mu=0.4$} & \multicolumn{1}{c|}{$\mu=0.8$} & \multicolumn{1}{c|}{$\mu=1.6$} \\ \midrule
\multicolumn{1}{|l|}{$\delta=0$}  & \multicolumn{1}{|c|} {0.426} & \multicolumn{1}{|c|} {0.455} & \multicolumn{1}{|c|} {0.279} & \multicolumn{1}{|c|} {0.244} & \multicolumn{1}{|c|} {0.228}                                \\ \midrule
\multicolumn{1}{|l|}{$\delta=0.1$}  & \multicolumn{1}{|c|} {0.274} & \multicolumn{1}{|c|} {0.447} & \multicolumn{1}{|c|} {0.284} & \multicolumn{1}{|c|} {0.243} & \multicolumn{1}{|c|} {0.235}                                \\ \midrule
\multicolumn{1}{|l|}{$\delta=0.2$}  & \multicolumn{1}{|c|} {0.243} & \multicolumn{1}{|c|} {0.425} & \multicolumn{1}{|c|} {0.272} & \multicolumn{1}{|c|} {0.242} & \multicolumn{1}{|c|} {0.231}                              \\ \midrule
\multicolumn{1}{|l|}{$\delta=0.3$}  & \multicolumn{1}{|c|} {0.232} & \multicolumn{1}{|c|} {0.443} & \multicolumn{1}{|c|} {0.286} & \multicolumn{1}{|c|} {0.246} & \multicolumn{1}{|c|} {0.23}                                \\ \bottomrule
\end{tabular}
\end{table}

\begin{table}[!htb]
\caption{Expected values of effort levels (authors) computed for $\epsilon=0.1$ over last 3k rounds.}
\label{table:expected_values_eps_0.1_authors_last_3000}
\centering
\begin{tabular}{@{}l|l|l|l|l|l|@{}}
\cmidrule(l){2-6}
                     & Double Blind & \multicolumn{4}{c|}{Open Review}  \\ \cmidrule(l){1-6} 
\multicolumn{1}{|c|}{\diagbox{$\delta$}{$\mu$}}                  & \multicolumn{1}{c|}{$\mu=0$} & \multicolumn{1}{c|}{$\mu=0.2$} & \multicolumn{1}{c|}{$\mu=0.4$} & \multicolumn{1}{c|}{$\mu=0.8$} & \multicolumn{1}{c|}{$\mu=1.6$} \\ \midrule
\multicolumn{1}{|l|}{$\delta=0$}  & \multicolumn{1}{|c|} {0.702} & \multicolumn{1}{|c|} {0.702} & \multicolumn{1}{|c|} {0.698} & \multicolumn{1}{|c|} {0.65} & \multicolumn{1}{|c|} {0.649}                                \\ \midrule
\multicolumn{1}{|l|}{$\delta=0.1$}  & \multicolumn{1}{|c|} {0.691} & \multicolumn{1}{|c|} {0.714} & \multicolumn{1}{|c|} {0.684} & \multicolumn{1}{|c|} {0.658} & \multicolumn{1}{|c|} {0.616}                                \\ \midrule
\multicolumn{1}{|l|}{$\delta=0.2$}  & \multicolumn{1}{|c|} {0.652} & \multicolumn{1}{|c|} {0.715} & \multicolumn{1}{|c|} {0.69} & \multicolumn{1}{|c|} {0.645} & \multicolumn{1}{|c|} {0.65}                              \\ \midrule
\multicolumn{1}{|l|}{$\delta=0.3$}  & \multicolumn{1}{|c|} {0.639} & \multicolumn{1}{|c|} {0.716} & \multicolumn{1}{|c|} {0.688} & \multicolumn{1}{|c|} {0.67} & \multicolumn{1}{|c|} {0.659}                                \\ \bottomrule
\end{tabular}
\end{table}

\begin{table}[!htb]
\caption{Expected values of threshold levels (reviewers) computed for $\epsilon=0.1$ over last 3k rounds.}
\label{table:expected_values_eps_0.1_reviewers_last_3000}
\centering
\begin{tabular}{@{}l|l|l|l|l|l|@{}}
\cmidrule(l){2-6}
                     & Double Blind & \multicolumn{4}{c|}{Open Review}  \\ \cmidrule(l){1-6} 
\multicolumn{1}{|c|}{\diagbox{$\delta$}{$\mu$}}                  & \multicolumn{1}{c|}{$\mu=0$} & \multicolumn{1}{c|}{$\mu=0.2$} & \multicolumn{1}{c|}{$\mu=0.4$} & \multicolumn{1}{c|}{$\mu=0.8$} & \multicolumn{1}{c|}{$\mu=1.6$} \\ \midrule
\multicolumn{1}{|l|}{$\delta=0$}  & \multicolumn{1}{|c|} {0.439} & \multicolumn{1}{|c|} {0.445} & \multicolumn{1}{|c|} {0.286} & \multicolumn{1}{|c|} {0.251} & \multicolumn{1}{|c|} {0.224}                                \\ \midrule
\multicolumn{1}{|l|}{$\delta=0.1$}  & \multicolumn{1}{|c|} {0.272} & \multicolumn{1}{|c|} {0.458} & \multicolumn{1}{|c|} {0.285} & \multicolumn{1}{|c|} {0.242} & \multicolumn{1}{|c|} {0.23}                                \\ \midrule
\multicolumn{1}{|l|}{$\delta=0.2$}  & \multicolumn{1}{|c|} {0.244} & \multicolumn{1}{|c|} {0.435} & \multicolumn{1}{|c|} {0.269} & \multicolumn{1}{|c|} {0.234} & \multicolumn{1}{|c|} {0.228}                              \\ \midrule
\multicolumn{1}{|l|}{$\delta=0.3$}  & \multicolumn{1}{|c|} {0.228} & \multicolumn{1}{|c|} {0.46} & \multicolumn{1}{|c|} {0.269} & \multicolumn{1}{|c|} {0.239} & \multicolumn{1}{|c|} {0.228}                                \\ \bottomrule
\end{tabular}
\end{table}
  
\begin{table}[!htb]
\caption{Expected values of effort levels (authors) computed for $\epsilon=0.2$ over an entire round of simulation (13k rounds).}
\label{table:expected_values_eps_0.2_authors_all}
\centering
\begin{tabular}{@{}l|l|l|l|l|l|@{}}
\cmidrule(l){2-6}
                     & Double Blind & \multicolumn{4}{c|}{Open Review}  \\ \cmidrule(l){1-6} 
\multicolumn{1}{|c|}{\diagbox{$\delta$}{$\mu$}}                  & \multicolumn{1}{c|}{$\mu=0$} & \multicolumn{1}{c|}{$\mu=0.2$} & \multicolumn{1}{c|}{$\mu=0.4$} & \multicolumn{1}{c|}{$\mu=0.8$} & \multicolumn{1}{c|}{$\mu=1.6$} \\ \midrule
\multicolumn{1}{|l|}{$\delta=0$}  & \multicolumn{1}{|c|} {0.765} & \multicolumn{1}{|c|} {0.772} & \multicolumn{1}{|c|} {0.733} & \multicolumn{1}{|c|} {0.758} & \multicolumn{1}{|c|} {0.733}                                \\ \midrule
\multicolumn{1}{|l|}{$\delta=0.1$}  & \multicolumn{1}{|c|} {0.723} & \multicolumn{1}{|c|} {0.733} & \multicolumn{1}{|c|} {0.73} & \multicolumn{1}{|c|} {0.729} & \multicolumn{1}{|c|} {0.736}                                \\ \midrule
\multicolumn{1}{|l|}{$\delta=0.2$}  & \multicolumn{1}{|c|} {0.703} & \multicolumn{1}{|c|} {0.703} & \multicolumn{1}{|c|} {0.721} & \multicolumn{1}{|c|} {0.729} & \multicolumn{1}{|c|} {0.732}                              \\ \midrule
\multicolumn{1}{|l|}{$\delta=0.3$}  & \multicolumn{1}{|c|} {0.711} & \multicolumn{1}{|c|} {0.71} & \multicolumn{1}{|c|} {0.704} & \multicolumn{1}{|c|} {0.707} & \multicolumn{1}{|c|} {0.706}                                \\ \bottomrule
\end{tabular}
\end{table}

\begin{table}[!htb]
\caption{Expected values of threshold levels (reviewers) computed for $\epsilon=0.2$ over an entire round of simulation (13k rounds).}
\label{table:expected_values_eps_0.2_reviewers_all}
\centering
\begin{tabular}{@{}l|l|l|l|l|l|@{}}
\cmidrule(l){2-6}
                     & Double Blind & \multicolumn{4}{c|}{Open Review}  \\ \cmidrule(l){1-6} 
\multicolumn{1}{|c|}{\diagbox{$\delta$}{$\mu$}}                  & \multicolumn{1}{c|}{$\mu=0$} & \multicolumn{1}{c|}{$\mu=0.2$} & \multicolumn{1}{c|}{$\mu=0.4$} & \multicolumn{1}{c|}{$\mu=0.8$} & \multicolumn{1}{c|}{$\mu=1.6$} \\ \midrule
\multicolumn{1}{|l|}{$\delta=0$}  & \multicolumn{1}{|c|} {0.45} & \multicolumn{1}{|c|} {0.455} & \multicolumn{1}{|c|} {0.44} & \multicolumn{1}{|c|} {0.427} & \multicolumn{1}{|c|} {0.425}                                \\ \midrule
\multicolumn{1}{|l|}{$\delta=0.1$}  & \multicolumn{1}{|c|} {0.28} & \multicolumn{1}{|c|} {0.29} & \multicolumn{1}{|c|} {0.276} & \multicolumn{1}{|c|} {0.283} & \multicolumn{1}{|c|} {0.285}                                \\ \midrule
\multicolumn{1}{|l|}{$\delta=0.2$}  & \multicolumn{1}{|c|} {0.248} & \multicolumn{1}{|c|} {0.242} & \multicolumn{1}{|c|} {0.239} & \multicolumn{1}{|c|} {0.24} & \multicolumn{1}{|c|} {0.241}                              \\ \midrule
\multicolumn{1}{|l|}{$\delta=0.3$}  & \multicolumn{1}{|c|} {0.23} & \multicolumn{1}{|c|} {0.232} & \multicolumn{1}{|c|} {0.229} & \multicolumn{1}{|c|} {0.229} & \multicolumn{1}{|c|} {0.231}                                \\ \bottomrule
\end{tabular}
\end{table}

\begin{table}[!htb]
\caption{Expected values of effort levels (authors) computed for $\epsilon=0.2$ over last 3k rounds.}
\label{table:expected_values_eps_0.2_authors_last_3000}
\centering
\begin{tabular}{@{}l|l|l|l|l|l|@{}}
\cmidrule(l){2-6}
                     & Double Blind & \multicolumn{4}{c|}{Open Review}  \\ \cmidrule(l){1-6} 
\multicolumn{1}{|c|}{\diagbox{$\delta$}{$\mu$}}                  & \multicolumn{1}{c|}{$\mu=0$} & \multicolumn{1}{c|}{$\mu=0.2$} & \multicolumn{1}{c|}{$\mu=0.4$} & \multicolumn{1}{c|}{$\mu=0.8$} & \multicolumn{1}{c|}{$\mu=1.6$} \\ \midrule
\multicolumn{1}{|l|}{$\delta=0$}  & \multicolumn{1}{|c|} {0.767} & \multicolumn{1}{|c|} {0.778} & \multicolumn{1}{|c|} {0.714} & \multicolumn{1}{|c|} {0.765} & \multicolumn{1}{|c|} {0.726}                                \\ \midrule
\multicolumn{1}{|l|}{$\delta=0.1$}  & \multicolumn{1}{|c|} {0.721} & \multicolumn{1}{|c|} {0.712} & \multicolumn{1}{|c|} {0.749} & \multicolumn{1}{|c|} {0.719} & \multicolumn{1}{|c|} {0.738}                                \\ \midrule
\multicolumn{1}{|l|}{$\delta=0.2$}  & \multicolumn{1}{|c|} {0.7} & \multicolumn{1}{|c|} {0.718} & \multicolumn{1}{|c|} {0.727} & \multicolumn{1}{|c|} {0.735} & \multicolumn{1}{|c|} {0.739}                              \\ \midrule
\multicolumn{1}{|l|}{$\delta=0.3$}  & \multicolumn{1}{|c|} {0.711} & \multicolumn{1}{|c|} {0.719} & \multicolumn{1}{|c|} {0.701} & \multicolumn{1}{|c|} {0.706} & \multicolumn{1}{|c|} {0.689}                                \\ \bottomrule
\end{tabular}
\end{table}

\begin{table}[!htb]
\caption{Expected values of threshold levels (reviewers) computed for $\epsilon=0.2$ over last 3k rounds.}
\label{table:expected_values_eps_0.2_reviewers_last_3000}
\centering
\begin{tabular}{@{}l|l|l|l|l|l|@{}}
\cmidrule(l){2-6}
                     & Double Blind & \multicolumn{4}{c|}{Open Review}  \\ \cmidrule(l){1-6} 
\multicolumn{1}{|c|}{\diagbox{$\delta$}{$\mu$}}                  & \multicolumn{1}{c|}{$\mu=0$} & \multicolumn{1}{c|}{$\mu=0.2$} & \multicolumn{1}{c|}{$\mu=0.4$} & \multicolumn{1}{c|}{$\mu=0.8$} & \multicolumn{1}{c|}{$\mu=1.6$} \\ \midrule
\multicolumn{1}{|l|}{$\delta=0$}  & \multicolumn{1}{|c|} {0.448} & \multicolumn{1}{|c|} {0.433} & \multicolumn{1}{|c|} {0.42} & \multicolumn{1}{|c|} {0.41} & \multicolumn{1}{|c|} {0.488}                                \\ \midrule
\multicolumn{1}{|l|}{$\delta=0.1$}  & \multicolumn{1}{|c|} {0.261} & \multicolumn{1}{|c|} {0.28} & \multicolumn{1}{|c|} {0.259} & \multicolumn{1}{|c|} {0.281} & \multicolumn{1}{|c|} {0.262}                                \\ \midrule
\multicolumn{1}{|l|}{$\delta=0.2$}  & \multicolumn{1}{|c|} {0.236} & \multicolumn{1}{|c|} {0.24} & \multicolumn{1}{|c|} {0.233} & \multicolumn{1}{|c|} {0.238} & \multicolumn{1}{|c|} {0.239}                              \\ \midrule
\multicolumn{1}{|l|}{$\delta=0.3$}  & \multicolumn{1}{|c|} {0.226} & \multicolumn{1}{|c|} {0.228} & \multicolumn{1}{|c|} {0.224} & \multicolumn{1}{|c|} {0.227} & \multicolumn{1}{|c|} {0.226}                                \\ \bottomrule
\end{tabular}
\end{table}
  
\begin{table}[!htb]
\caption{Expected values of effort levels (authors) computed for $\epsilon=0.3$ over an entire round of simulation (13k rounds).}
\label{table:expected_values_eps_0.3_authors_all}
\centering
\begin{tabular}{@{}l|l|l|l|l|l|@{}}
\cmidrule(l){2-6}
                     & Double Blind & \multicolumn{4}{c|}{Open Review}  \\ \cmidrule(l){1-6} 
\multicolumn{1}{|c|}{\diagbox{$\delta$}{$\mu$}}                  & \multicolumn{1}{c|}{$\mu=0$} & \multicolumn{1}{c|}{$\mu=0.2$} & \multicolumn{1}{c|}{$\mu=0.4$} & \multicolumn{1}{c|}{$\mu=0.8$} & \multicolumn{1}{c|}{$\mu=1.6$} \\ \midrule
\multicolumn{1}{|l|}{$\delta=0$}  & \multicolumn{1}{|c|} {{0.782}} & \multicolumn{1}{|c|} {0.774} & \multicolumn{1}{|c|} {0.777} & \multicolumn{1}{|c|} {0.78} & \multicolumn{1}{|c|} {0.781}                                \\ \midrule
\multicolumn{1}{|l|}{$\delta=0.1$}  & \multicolumn{1}{|c|} {0.768} & \multicolumn{1}{|c|} {{0.772}} & \multicolumn{1}{|c|} {0.768} & \multicolumn{1}{|c|} {{0.772}} & \multicolumn{1}{|c|} {0.77}                                \\ \midrule
\multicolumn{1}{|l|}{$\delta=0.2$}  & \multicolumn{1}{|c|} {0.758} & \multicolumn{1}{|c|} {{0.766}} & \multicolumn{1}{|c|} {0.764} & \multicolumn{1}{|c|} {0.762} & \multicolumn{1}{|c|} {0.759}                              \\ \midrule
\multicolumn{1}{|l|}{$\delta=0.3$}  & \multicolumn{1}{|c|} {0.75} & \multicolumn{1}{|c|} {{0.758}} & \multicolumn{1}{|c|} {0.752} & \multicolumn{1}{|c|} {0.756} & \multicolumn{1}{|c|} {0.755}                                \\ \bottomrule
\end{tabular}
\end{table}

\begin{table}[!htb]
\caption{Expected values of threshold levels (reviewers) computed for $\epsilon=0.3$ over an entire round of simulation (13k rounds).}
\label{table:expected_values_eps_0.3_reviewers_all}
\centering
\begin{tabular}{@{}l|l|l|l|l|l|@{}}
\cmidrule(l){2-6}
                     & Double Blind & \multicolumn{4}{c|}{Open Review}  \\ \cmidrule(l){1-6} 
\multicolumn{1}{|c|}{\diagbox{$\delta$}{$\mu$}}                  & \multicolumn{1}{c|}{$\mu=0$} & \multicolumn{1}{c|}{$\mu=0.2$} & \multicolumn{1}{c|}{$\mu=0.4$} & \multicolumn{1}{c|}{$\mu=0.8$} & \multicolumn{1}{c|}{$\mu=1.6$} \\ \midrule
\multicolumn{1}{|l|}{$\delta=0$}  & \multicolumn{1}{|c|} {0.462} & \multicolumn{1}{|c|} {{0.471}} & \multicolumn{1}{|c|} {0.44} & \multicolumn{1}{|c|} {0.417} & \multicolumn{1}{|c|} {0.441}                                \\ \midrule
\multicolumn{1}{|l|}{$\delta=0.1$}  & \multicolumn{1}{|c|} {0.272} & \multicolumn{1}{|c|} {0.278} & \multicolumn{1}{|c|} {{0.283}} & \multicolumn{1}{|c|} {0.281} & \multicolumn{1}{|c|} {0.28}                                \\ \midrule
\multicolumn{1}{|l|}{$\delta=0.2$}  & \multicolumn{1}{|c|} {{0.245}} & \multicolumn{1}{|c|} {{0.245}} & \multicolumn{1}{|c|} {0.241} & \multicolumn{1}{|c|} {{0.245}} & \multicolumn{1}{|c|} {0.242}                              \\ \midrule
\multicolumn{1}{|l|}{$\delta=0.3$}  & \multicolumn{1}{|c|} {{0.232}} & \multicolumn{1}{|c|} {{0.232}} & \multicolumn{1}{|c|} {0.231} & \multicolumn{1}{|c|} {0.23} & \multicolumn{1}{|c|} {0.23}                                \\ \bottomrule
\end{tabular}
\end{table}

\begin{table}[!htb]
\caption{Expected values of effort levels (authors) computed for $\epsilon=0.3$ over last 3k rounds.}
\label{table:expected_values_eps_0.3_authors_last_3000}
\centering
\begin{tabular}{@{}l|l|l|l|l|l|@{}}
\cmidrule(l){2-6}
                     & Double Blind & \multicolumn{4}{c|}{Open Review}  \\ \cmidrule(l){1-6} 
\multicolumn{1}{|c|}{\diagbox{$\delta$}{$\mu$}}                  & \multicolumn{1}{c|}{$\mu=0$} & \multicolumn{1}{c|}{$\mu=0.2$} & \multicolumn{1}{c|}{$\mu=0.4$} & \multicolumn{1}{c|}{$\mu=0.8$} & \multicolumn{1}{c|}{$\mu=1.6$} \\ \midrule
\multicolumn{1}{|l|}{$\delta=0$}  & \multicolumn{1}{|c|} {0.785} & \multicolumn{1}{|c|} {0.779} & \multicolumn{1}{|c|} {0.781} & \multicolumn{1}{|c|} {{0.789}} & \multicolumn{1}{|c|} {0.787}                                \\ \midrule
\multicolumn{1}{|l|}{$\delta=0.1$}  & \multicolumn{1}{|c|} {0.774} & \multicolumn{1}{|c|} {0.773} & \multicolumn{1}{|c|} {0.772} & \multicolumn{1}{|c|} {0.774} & \multicolumn{1}{|c|} {{0.778}}                                \\ \midrule
\multicolumn{1}{|l|}{$\delta=0.2$}  & \multicolumn{1}{|c|} {0.765} & \multicolumn{1}{|c|} {0.763} & \multicolumn{1}{|c|} {0.763} & \multicolumn{1}{|c|} {{0.768}} & \multicolumn{1}{|c|} {0.764}                              \\ \midrule
\multicolumn{1}{|l|}{$\delta=0.3$}  & \multicolumn{1}{|c|} {0.747} & \multicolumn{1}{|c|} {0.756} & \multicolumn{1}{|c|} {0.758} & \multicolumn{1}{|c|} {{0.762}} & \multicolumn{1}{|c|} {0.751}                                \\ \bottomrule
\end{tabular}
\end{table}

\begin{table}[!htb]
\caption{Expected values of threshold levels (reviewers) computed for $\epsilon=0.3$ over last 3k rounds.}
\label{table:expected_values_eps_0.3_reviewers_last_3000}
\centering
\begin{tabular}{@{}l|l|l|l|l|l|@{}}
\cmidrule(l){2-6}
                     & Double Blind & \multicolumn{4}{c|}{Open Review}  \\ \cmidrule(l){1-6} 
\multicolumn{1}{|c|}{\diagbox{$\delta$}{$\mu$}}                  & \multicolumn{1}{c|}{$\mu=0$} & \multicolumn{1}{c|}{$\mu=0.2$} & \multicolumn{1}{c|}{$\mu=0.4$} & \multicolumn{1}{c|}{$\mu=0.8$} & \multicolumn{1}{c|}{$\mu=1.6$} \\ \midrule
\multicolumn{1}{|l|}{$\delta=0$}  & \multicolumn{1}{|c|} {0.445} & \multicolumn{1}{|c|} {0.424} & \multicolumn{1}{|c|} {{0.472}} & \multicolumn{1}{|c|} {0.456} & \multicolumn{1}{|c|} {0.391}                                \\ \midrule
\multicolumn{1}{|l|}{$\delta=0.1$}  & \multicolumn{1}{|c|} {0.266} & \multicolumn{1}{|c|} {0.265} & \multicolumn{1}{|c|} {{0.29}} & \multicolumn{1}{|c|} {0.257} & \multicolumn{1}{|c|} {0.268}                                \\ \midrule
\multicolumn{1}{|l|}{$\delta=0.2$}  & \multicolumn{1}{|c|} {{0.243}} & \multicolumn{1}{|c|} {0.235} & \multicolumn{1}{|c|} {0.234} & \multicolumn{1}{|c|} {0.239} & \multicolumn{1}{|c|} {0.235}                              \\ \midrule
\multicolumn{1}{|l|}{$\delta=0.3$}  & \multicolumn{1}{|c|} {0.226} & \multicolumn{1}{|c|} {{0.229}} & \multicolumn{1}{|c|} {0.226} & \multicolumn{1}{|c|} {0.226} & \multicolumn{1}{|c|} {0.225}                                \\ \bottomrule
\end{tabular}
\end{table}

\begin{table}[!htb]
\caption{Expected values of effort levels (authors) computed for $\epsilon=0.4$ over an entire round of simulation (13k rounds).}
\label{table:expected_values_eps_0.4_authors_all}
\centering
\begin{tabular}{@{}l|l|l|l|l|l|@{}}
\cmidrule(l){2-6}
                     & Double Blind & \multicolumn{4}{c|}{Open Review}  \\ \cmidrule(l){1-6} 
\multicolumn{1}{|c|}{\diagbox{$\delta$}{$\mu$}}                  & \multicolumn{1}{c|}{$\mu=0$} & \multicolumn{1}{c|}{$\mu=0.2$} & \multicolumn{1}{c|}{$\mu=0.4$} & \multicolumn{1}{c|}{$\mu=0.8$} & \multicolumn{1}{c|}{$\mu=1.6$} \\ \midrule
\multicolumn{1}{|l|}{$\delta=0$}  & \multicolumn{1}{|c|} {{0.786}} & \multicolumn{1}{|c|} {0.785} & \multicolumn{1}{|c|} {0.779} & \multicolumn{1}{|c|} {0.775} & \multicolumn{1}{|c|} {0.77}                                \\ \midrule
\multicolumn{1}{|l|}{$\delta=0.1$}  & \multicolumn{1}{|c|} {0.778} & \multicolumn{1}{|c|} {{0.787}} & \multicolumn{1}{|c|} {0.777} & \multicolumn{1}{|c|} {0.773} & \multicolumn{1}{|c|} {0.771}                                \\ \midrule
\multicolumn{1}{|l|}{$\delta=0.2$}  & \multicolumn{1}{|c|} {0.775} & \multicolumn{1}{|c|} {{0.787}} & \multicolumn{1}{|c|} {0.78} & \multicolumn{1}{|c|} {0.774} & \multicolumn{1}{|c|} {0.773}                              \\ \midrule
\multicolumn{1}{|l|}{$\delta=0.3$}  & \multicolumn{1}{|c|} {0.77} & \multicolumn{1}{|c|} {{0.787}} & \multicolumn{1}{|c|} {0.777} & \multicolumn{1}{|c|} {0.777} & \multicolumn{1}{|c|} {0.773}                                \\ \bottomrule
\end{tabular}
\end{table}

\begin{table}[!htb]
\caption{Expected values of threshold levels (reviewers) computed for $\epsilon=0.4$ over an entire round of simulation (13k rounds).}
\label{table:expected_values_eps_0.4_reviewers_all}
\centering
\begin{tabular}{@{}l|l|l|l|l|l|@{}}
\cmidrule(l){2-6}
                     & Double Blind & \multicolumn{4}{c|}{Open Review}  \\ \cmidrule(l){1-6} 
\multicolumn{1}{|c|}{\diagbox{$\delta$}{$\mu$}}                  & \multicolumn{1}{c|}{$\mu=0$} & \multicolumn{1}{c|}{$\mu=0.2$} & \multicolumn{1}{c|}{$\mu=0.4$} & \multicolumn{1}{c|}{$\mu=0.8$} & \multicolumn{1}{c|}{$\mu=1.6$} \\ \midrule
\multicolumn{1}{|l|}{$\delta=0$}  & \multicolumn{1}{|c|} {{0.466}} & \multicolumn{1}{|c|} {0.458} & \multicolumn{1}{|c|} {0.429} & \multicolumn{1}{|c|} {0.246} & \multicolumn{1}{|c|} {0.23}                                \\ \midrule
\multicolumn{1}{|l|}{$\delta=0.1$}  & \multicolumn{1}{|c|} {0.281} & \multicolumn{1}{|c|} {{0.449}} & \multicolumn{1}{|c|} {0.272} & \multicolumn{1}{|c|} {0.244} & \multicolumn{1}{|c|} {0.229}                                \\ \midrule
\multicolumn{1}{|l|}{$\delta=0.2$}  & \multicolumn{1}{|c|} {0.242} & \multicolumn{1}{|c|} {{0.442}} & \multicolumn{1}{|c|} {0.274} & \multicolumn{1}{|c|} {0.241} & \multicolumn{1}{|c|} {0.228}                              \\ \midrule
\multicolumn{1}{|l|}{$\delta=0.3$}  & \multicolumn{1}{|c|} {0.229} & \multicolumn{1}{|c|} {{0.429}} & \multicolumn{1}{|c|} {0.273} & \multicolumn{1}{|c|} {0.241} & \multicolumn{1}{|c|} {0.229}                                \\ \bottomrule
\end{tabular}
\end{table}

\begin{table}[!htb]
\caption{Expected values of effort levels (authors) computed for $\epsilon=0.4$ over last 3k rounds.}
\label{table:expected_values_eps_0.4_authors_last_3000}
\centering
\begin{tabular}{@{}l|l|l|l|l|l|@{}}
\cmidrule(l){2-6}
                     & Double Blind & \multicolumn{4}{c|}{Open Review}  \\ \cmidrule(l){1-6} 
\multicolumn{1}{|c|}{\diagbox{$\delta$}{$\mu$}}                  & \multicolumn{1}{c|}{$\mu=0$} & \multicolumn{1}{c|}{$\mu=0.2$} & \multicolumn{1}{c|}{$\mu=0.4$} & \multicolumn{1}{c|}{$\mu=0.8$} & \multicolumn{1}{c|}{$\mu=1.6$} \\ \midrule
\multicolumn{1}{|l|}{$\delta=0$}  & \multicolumn{1}{|c|} {{0.791}} & \multicolumn{1}{|c|} {0.786} & \multicolumn{1}{|c|} {0.776} & \multicolumn{1}{|c|} {0.777} & \multicolumn{1}{|c|} {0.769}                                \\ \midrule
\multicolumn{1}{|l|}{$\delta=0.1$}  & \multicolumn{1}{|c|} {0.778} & \multicolumn{1}{|c|} {{0.786}} & \multicolumn{1}{|c|} {0.778} & \multicolumn{1}{|c|} {0.777} & \multicolumn{1}{|c|} {0.776}                                \\ \midrule
\multicolumn{1}{|l|}{$\delta=0.2$}  & \multicolumn{1}{|c|} {0.779} & \multicolumn{1}{|c|} {{0.788}} & \multicolumn{1}{|c|} {0.782} & \multicolumn{1}{|c|} {0.774} & \multicolumn{1}{|c|} {0.77}                              \\ \midrule
\multicolumn{1}{|l|}{$\delta=0.3$}  & \multicolumn{1}{|c|} {0.774} & \multicolumn{1}{|c|} {{0.787}} & \multicolumn{1}{|c|} {0.778} & \multicolumn{1}{|c|} {0.777} & \multicolumn{1}{|c|} {0.773}                                \\ \bottomrule
\end{tabular}
\end{table}

\begin{table}[!htb]
\caption{Expected values of threshold levels (reviewers) computed for $\epsilon=0.4$ over last 3k rounds.}
\label{table:expected_values_eps_0.4_reviewers_last_3000}
\centering
\begin{tabular}{@{}l|l|l|l|l|l|@{}}
\cmidrule(l){2-6}
                     & Double Blind & \multicolumn{4}{c|}{Open Review}  \\ \cmidrule(l){1-6} 
\multicolumn{1}{|c|}{\diagbox{$\delta$}{$\mu$}}                  & \multicolumn{1}{c|}{$\mu=0$} & \multicolumn{1}{c|}{$\mu=0.2$} & \multicolumn{1}{c|}{$\mu=0.4$} & \multicolumn{1}{c|}{$\mu=0.8$} & \multicolumn{1}{c|}{$\mu=1.6$} \\ \midrule
\multicolumn{1}{|l|}{$\delta=0$}  & \multicolumn{1}{|c|} {{0.465}} & \multicolumn{1}{|c|} {0.457} & \multicolumn{1}{|c|} {0.277} & \multicolumn{1}{|c|} {0.245} & \multicolumn{1}{|c|} {0.224}                                \\ \midrule
\multicolumn{1}{|l|}{$\delta=0.1$}  & \multicolumn{1}{|c|} {0.263} & \multicolumn{1}{|c|} {{0.455}} & \multicolumn{1}{|c|} {0.265} & \multicolumn{1}{|c|} {0.235} & \multicolumn{1}{|c|} {0.226}                                \\ \midrule
\multicolumn{1}{|l|}{$\delta=0.2$}  & \multicolumn{1}{|c|} {0.243} & \multicolumn{1}{|c|} {{0.495}} & \multicolumn{1}{|c|} {0.27} & \multicolumn{1}{|c|} {0.233} & \multicolumn{1}{|c|} {0.223}                              \\ \midrule
\multicolumn{1}{|l|}{$\delta=0.3$}  & \multicolumn{1}{|c|} {0.224} & \multicolumn{1}{|c|} {{0.432}} & \multicolumn{1}{|c|} {0.257} & \multicolumn{1}{|c|} {0.233} & \multicolumn{1}{|c|} {0.224}                                \\ \bottomrule
\end{tabular}
\end{table}

\end{document}